\documentclass[conference]{IEEEtran}
\usepackage[firstpage]{draftwatermark}
\SetWatermarkText{\hspace*{6.2in}\raisebox{7in}
  {\includegraphics[scale=0.1]{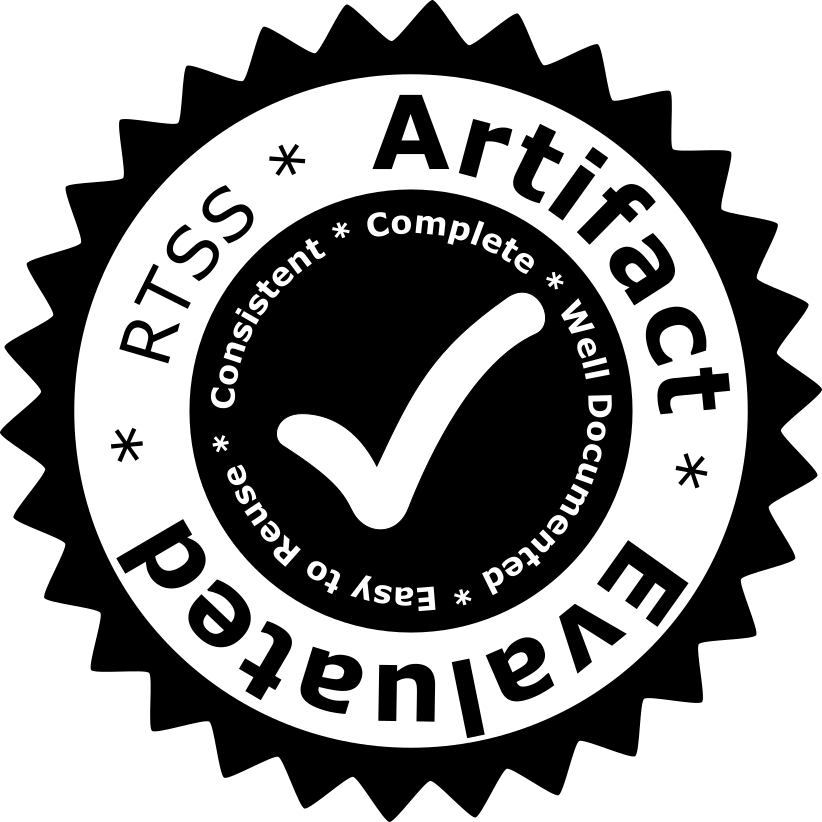}}}
\SetWatermarkAngle{0}

\IEEEoverridecommandlockouts
\usepackage{cite}
\usepackage{amsmath,amssymb,amsfonts}
\usepackage{algorithm}
\usepackage[noend]{algorithmic}
\usepackage{graphicx}
\usepackage{textcomp}
\usepackage{xcolor}
\usepackage{amsthm}
\usepackage{enumerate}
\usepackage{scalerel}
\usepackage{array}
\usepackage[hidelinks]{hyperref}
\usepackage{subfigure}
\usepackage{bbm}
\usepackage{dsfont}
\usepackage{url}

\def\BibTeX{{\rm B\kern-.05em{\sc i\kern-.025em b}\kern-.08em
    T\kern-.1667em\lower.7ex\hbox{E}\kern-.125emX}}

\newtheorem{theorem}{Theorem}
\newtheorem{lemma}{Lemma}
\newtheorem{proposition}{Proposition}
\newtheorem{definition}{Definition}

\DeclareMathOperator*{\argmin}{\arg\min}

\DeclareRobustCommand*{\IEEEauthorrefmark}[1]{%
  \raisebox{0pt}[0pt][0pt]{\textsuperscript{\footnotesize #1}}%
}
    
\begin{document}

\title{Energy-Efficient Real-Time Job Mapping and Resource Management in Mobile-Edge Computing
\thanks{This work was supported in part by the MoE Tier-2 grant MOE-T2EP20221-0006.}
}

\author{\IEEEauthorblockN{Chuanchao Gao\IEEEauthorrefmark{1,2}
, Niraj Kumar\IEEEauthorrefmark{1}, Arvind Easwaran\IEEEauthorrefmark{1,2}}
\IEEEauthorblockA{\IEEEauthorrefmark{1}\textit{College of Computing and Data Science}}
\IEEEauthorblockA{\IEEEauthorrefmark{2}\textit{Energy Research Institute @ NTU, Interdisciplinary Graduate Programme}}
\IEEEauthorblockA{
\textit{Nanyang Technological University, Singapore}\\
gaoc0008@e.ntu.edu.sg, niraj.kumar@ntu.edu.sg, arvinde@ntu.edu.sg}
}

\maketitle

\begin{abstract}
Mobile-edge computing (MEC) has emerged as a promising paradigm for enabling Internet of Things (IoT) devices to handle computation-intensive jobs. Due to the imperfect parallelization of algorithms for job processing on servers and the impact of IoT device mobility on data communication quality in wireless networks, it is crucial to jointly consider server resource allocation and IoT device mobility during job scheduling to fully benefit from MEC, which is often overlooked in existing studies. By jointly considering job scheduling, server resource allocation, and IoT device mobility, we investigate the deadline-constrained job offloading and resource management problem in MEC with both communication and computation contentions, aiming to maximize the total energy saved for IoT devices. For the offline version of the problem, where job information is known in advance, we formulate it as an Integer Linear Programming problem and propose an approximation algorithm, $\mathtt{LHJS}$, with a constant performance guarantee. For the online version, where job information is only known upon release, we propose a heuristic algorithm, $\mathtt{LBS}$, that is invoked whenever a job is released. Finally, we conduct experiments with parameters from real-world applications to evaluate their performance.

\end{abstract}


\section{Introduction}\label{sec:intro}
Internet of Things (IoT) has emerged as a forefront technology (Cisco predicts 500 billion IoT devices by 2030 \cite{ji2021guest}) driven by advancements in hardware, software, and communication technologies \cite{aazam2018fog} such as low-power wide-area networks, WiFi, and ultra-reliable low-latency communication of 5G.
Concurrently, the evolution of Artificial Intelligence has led to many computation-intensive IoT applications such as virtual/augmented reality \cite{schneider2017augmented}, image/video processing \cite{soyata2012cloud}, and object detection in autonomous driving \cite{yurtsever2020survey}. These applications, many with rigid timing constraints, pose significant challenges to IoT devices, which are typically battery-powered and resource-constrained for compactness and portability \cite{ma2011battery}.

Mobile-Edge Computing (MEC) has emerged as a promising paradigm enabling IoT devices to effectively support computation-intensive and time-critical applications.
In MEC, jobs are offloaded by end devices (EDs) to nearby access points (APs) through wireless networks and then forwarded to servers via a wired backhaul network for processing. Deploying servers close to EDs in MEC significantly reduces communication latency compared to cloud computing, enabling prompt responses to ED requests. Offloading computation-intensive jobs to servers conserves EDs' energy and accelerates job processing, but also introduces additional latency and energy consumption for job offloading. 
Furthermore, considering the limited communication and computation resources of APs and servers, respectively, efficient job mapping (to APs and servers) and resource management (for offloading, processing, and downloading) strategies in MEC become crucial, especially for time-critical applications. 


Resource management comprises two tasks: resource allocation and job scheduling. Resource allocation is concerned with the problem of how much resource to allocate to a job, i.e., in terms of amount of resource units (e.g., number of processing cores on servers).
Job scheduling, on the other hand, is concerned with the problem of when to schedule the allocated resources for the jobs so as to meet job deadlines.

In this study, we focus on the problem of job mapping and resource management for deadline-constrained jobs with the following considerations. For many applications, including those that utilize GPUs, resource utilization efficiency decreases as computation parallelism increases due to the imperfect parallelization of algorithms within an application. Always allocating full server resources to each job can lead to server resources being underutilized, highlighting the importance of considering resource allocations during scheduling. Orthogonally, the wireless network condition varies with ED mobility. Considering ED mobility during job scheduling can save energy for EDs, for example, by scheduling job offloading when the network condition is optimal between EDs and APs. Although some studies have explored the problem of job mapping and scheduling for deadline-constrained jobs in MEC~\cite{shuai2022transfer, ma2024latency, zhang2024computational, alameddine2019dynamic, sorkhoh2019workload, xu2023adaptive, lou2022cost, gao2023joint, abdi2024task, sang2024mobility, tiwari2024rate, zhu2018task, huang2023joint, huang2024mobility, meng2019dedas, han2020joint, pu2019chimera, pradhan2024towards, chauhan2024probabilistic, li2024uav, lai2024short}, to our best knowledge, the joint consideration of job mapping, resource management (including allocations) and ED mobility is absent.

This study addresses a deadline-constrained job mapping and resource management problem in MEC with both communication and computation contentions, aiming to maximize the total energy that can be saved for EDs; we refer to it as the Energy Maximization Job Scheduling Problem ($\mathsf{EMJS}$). Offloaded jobs compete with each other for both wireless bandwidth on APs and computation resource on servers. We jointly consider computation resource allocation on servers, job scheduling (for offloading, processing, and downloading) on APs and servers and ED mobility. Our model incorporates a versatile job mapping framework, enabling each job to be offloaded to any of its accessible APs and subsequently forwarded to one of the capable servers for processing. Upon completion, the job (result) is relayed back to any of its accessible APs and subsequently downloaded to its ED. Considering ED mobility, the accessible APs for offloading may differ from those for downloading. Additionally, each job can only be processed on servers possessing the resource type demanded by the job. Each server offers multiple predefined computation resource allocation options; each job can be allocated any one of these options resulting in different processing durations. Notably, in scenarios where MEC comprises only one AP and a co-located server with each job always being allocated full server resources, $\mathsf{EMJS}$ is equivalent to a three-machine flow shop problem \cite{ben2020maximizing}, known as NP-Hard. Hence, considering options for computation resource allocation and MECs with several APs and servers, the general $\mathsf{EMJS}$ is also NP-Hard.

We consider two versions of $\mathsf{EMJS}$ in this paper: online and offline. In online $\mathsf{EMJS}$, information about each job is only available upon its release in the system. In contrast, in offline $\mathsf{EMJS}$, information about all the jobs is available apriori. To address offline $\mathsf{EMJS}$, we first formulate it as an Integer Linear Programming (ILP) problem by enumerating all possible \emph{schedule instances}, where a schedule instance is a combination of (i) mapping of jobs to APs and servers, (ii) computation resource allocation on servers, and (iii) starting times for offloading, processing, and downloading (\emph{jobs are assumed to be scheduled non-preemptively on each resource}). Then, we present a pseudo-polynomial approximation algorithm, called Light-Heavy Job Scheduling ($\mathtt{LHJS}$), with a proven constant approximation ratio. For online $\mathsf{EMJS}$, we propose a heuristic algorithm, called Load Balanced Job Scheduling ($\mathtt{LBS}$), that is invoked whenever a job is released in the system. The contributions of this work are summarized below.
\begin{itemize}
    \item We address the $\mathsf{EMJS}$ in MEC with both communication and computation contentions, aiming to maximize the total saved energy for EDs. We jointly consider job mapping, resource management and ED mobility.
    \item We formulate the offline $\mathsf{EMJS}$ as an ILP problem by enumerating all schedule instances. Then, we propose, to the best of our knowledge, a first approximation algorithm ($\mathtt{LHJS}$) for $\mathsf{EMJS}$ with a constant approximation ratio. $\mathtt{LHJS}$ divides all instances into two sets based on computation resource allocation and schedules them separately.
    \item For the online version of $\mathsf{EMJS}$, we present a heuristic scheduling algorithm called $\mathtt{LBS}$ by scheduling each job on the least busy server and allocating the minimum feasible computation resource to each job.
    \item We experimentally evaluate $\mathtt{LHJS}$ and $\mathtt{LBS}$ using profiled data of real-world applications. Results show that $\mathtt{LHJS}$ outperforms its baseline algorithm \cite{zhu2018task} by $11.8\%$ in offline scheduling, and $\mathtt{LBS}$ achieves an average of $83.2\%$ of the optimal saved energy in online scheduling.
\end{itemize}


\begin{figure*}[t]
    \centering
    \subfigure[]{\includegraphics[width=0.25\textwidth, page=2]{figures/system.pdf} \label{fig:system1}} 
    \hfill
    \subfigure[]{\includegraphics[width=0.21\textwidth, page=1]{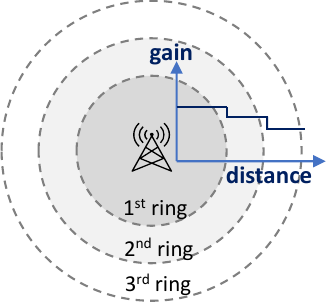} \label{fig:system2}}
    \hfill
    \subfigure[]{\includegraphics[width=0.32\textwidth, page=3]{figures/system.pdf} \label{fig:system3}}
    \caption{(a) a mobile edge computing system; (b) partition the coverage area of a vAP into three rings; (c) an example of an offloaded job. A job generated by a vehicle is offloaded to vAP $m_u$ when covered by network ring $m_{u1}$, forwarded to server $m_p$, processed on $m_p$, (result) forwarded to vAP $m_d$, and (result) downloaded from $m_d$ when covered by network ring $m_{d1}$.}
    \label{fig:system}
\end{figure*}

\section{Related Work}\label{sec:literature}
Due to the promising prospects of MEC, research interest in job mapping and resource management has grown significantly. Readers seeking a comprehensive overview of this topic can refer to related surveys \cite{jamil2022resource, luo2021resource, ramanathan2020survey}. Research problems can be categorized into deadline-constrained problems and deadline-free (or response time minimization) problems. This section reviews state-of-the-art research on \textit{deadline-constrained} job mapping and \textit{resource management} in MEC.

Depending on whether resource contention is taken into consideration (either computation or communication), these studies can be categorized into those (i) with no resource contention \cite{shuai2022transfer, ma2024latency, zhang2024computational}, (ii) only computation contention \cite{alameddine2019dynamic, sorkhoh2019workload, xu2023adaptive, lou2022cost, gao2023joint, abdi2024task, sang2024mobility, tiwari2024rate} or only communication contention \cite{zhu2018task, huang2023joint, huang2024mobility}, and (iii) both communication and computation contentions \cite{meng2019dedas, han2020joint, pu2019chimera, pradhan2024towards, chauhan2024probabilistic, li2024uav, lai2024short}. 

Some studies \cite{alameddine2019dynamic, xu2023adaptive, gao2023joint} have jointly considered job scheduling and computation resource allocation. These studies considered deploying services or virtual machines (VMs) on servers for job processing, where multiple services or VMs could run on the same server simultaneously, provided the total allocated resources did not exceed the server's resource capacity. Jobs were then mapped to and processed by each service or VM. Under this setup, all jobs mapped to the same service or VM are allocated the same amount of computation resource. Moreover, Gao \emph{et al.} \cite{gao2023joint} assigned equal amount of computation resources to a fixed number of VMs for each server. Unlike these studies, our research considers resource allocations for each job separately, i.e., a job mapped to a server can be allocated with any of the offered computation resource allocation options.
Furthermore, \textit{none of the above studies account for communication contention for job offloading/downloading or ED mobility}.

Some studies \cite{zhu2018task, sorkhoh2019workload, sang2024mobility, huang2024mobility} have considered ED mobility in MEC. These studies assumed that ED trajectories were predictable, allowing the determination of accessible APs for jobs and the corresponding network channel status. 
Sang \emph{et al.} \cite{sang2024mobility} assumed that channel noise was a function of the physical distance between EDs and APs. Sorkhoh \emph{et al.} \cite{sorkhoh2019workload} and Zhu \emph{et al.} \cite{zhu2018task} divided each wireless network into several ranges such that the channel gain (or data rate) within each range remained constant, thereby reducing the accuracy requirement for trajectory prediction; in our study, we consider a similar mobility model. However, unlike our work, \textit{none of these studies considered computation resource allocation during job scheduling}. Furthermore, Sorkhoh \emph{et al.} \cite{sorkhoh2019workload} and Sang \emph{et al.} \cite{sang2024mobility} did not consider communication contention, while Zhu \emph{et al.} \cite{zhu2018task} and Huang and Yu \cite{huang2024mobility} did not account for computation resource contention. Moreover, Zhu \emph{et al.} \cite{zhu2018task} proposed an offline scheduling algorithm with a parameterized approximation bound, which could be applied to our problem after some modifications (described in Section \ref{subsec:exp-setup}); thus, the modified algorithm of Zhu \emph{et al.} is used as the offline baseline algorithm in evaluating $\mathtt{LHJS}$ in our experiments.


Existing solutions to deadline-constrained job mapping and resource management problems can be classified into heuristic algorithms \cite{sorkhoh2019workload, shuai2022transfer, ma2024latency, zhang2024computational, xu2023adaptive, lou2022cost, gao2023joint, abdi2024task, sang2024mobility, tiwari2024rate, zhu2018task, huang2023joint, huang2024mobility, meng2019dedas, han2020joint, pu2019chimera, pradhan2024towards, chauhan2024probabilistic, li2024uav} and exponential-time exact algorithms \cite{alameddine2019dynamic, lai2024short}. Among heuristic algorithms, approximation algorithms \cite{gao2023joint, zhu2018task, meng2019dedas} offer additional theoretical guarantees. Unlike our work, Gao \emph{et al.} \cite{gao2023joint} did not consider communication contention, and Meng \emph{et al.} \cite{meng2019dedas} did not address the computation resource allocation during job scheduling and only derived parameterized approximation bounds. Furthermore, both studies did not consider ED mobility.


Our study jointly considers job mapping, resource management (including computation resource allocation) and ED mobility in MEC with both communication and computation contentions. Furthermore, we propose the first approximation algorithm with a constant approximation ratio for this problem.


\section{System Model and Problem Formulation}\label{sec:system}
\subsection{Mobile Edge Computing System Architecture}\label{subsec:architecture}
A MEC comprises EDs, APs, and servers (Fig. \ref{fig:system1}). 
A job can be offloaded by its ED to one of its accessible APs via a wireless network, and then forwarded to one of its capable servers (i.e., servers that possess the resource type demanded by the job) for processing through the backhaul network. After processing, the job (result) is relayed to one of its accessible APs and subsequently downloaded from the AP to its ED. Due to ED mobility, the set of accessible APs of a job may change over time so that APs used for offloading and downloading may be different. In this paper, \textit{we use discrete time, so that all the temporal parameters are specified as integers}. A notation summary is provided in Table \ref{table:notation}.

In a wireless network, the uplink and downlink channels of each AP are usually separated by a guard band, facilitating simultaneous data offloading and downloading within the same wireless network without interference. For analytical simplicity, we define two virtual APs, referred to as vAPs, corresponding to each AP such that each vAP exclusively contains one uplink or downlink channel. Let $\mathcal{M}^u$ be the set of $M^u$ vAPs with uplink channels, and $\mathcal{M}^d$ be the set of $M^d$ vAPs with downlink channels. Since the channel gain of wireless networks decreases with the increasing physical distance between EDs and vAPs \cite{zhu2018task}, for each vAP, we divide its wireless network coverage area into several \textit{rings} such that the channel gains remain unchanged (from a practical viewpoint) within the same ring (Fig.~\ref{fig:system2}); we use $m_{ir}$ to denote the $r$-th ring of vAP $m_i \in \mathcal{M}^u \cup \mathcal{M}^d$. In this paper, we consider heterogeneous servers that provide different types of computation resources, and a job can only be processed on a server possessing the resource type demanded by the job. Let $\mathcal{M}^p$ be the set of $M^p$ servers.

We refer to a vAP or server in MEC as a \textit{machine}, and let $\mathcal{M} = \mathcal{M}^u \cup \mathcal{M}^p \cup \mathcal{M}^d$ be the set of all machines in MEC, and let $M = \max\{M^u, M^p, M^d\}$. Let $\alpha_i$ denote the resource capacity of a machine $m_i \in \mathcal{M}$. The bandwidth capacity of a vAP is measured in MegaHertz (MHz). We consider Orthogonal Frequency Division Multiplexing (OFDM) as the network access method for EDs, which has been widely used in modern wireless networks such as WiFi-5 \cite{intel_wifi_protocal}, 4G-LTE \cite{hamdi2021tran}, and 5G \cite{ankarali2020enhanced}. In OFDM-enabled networks, each channel uses multiple overlapping but orthogonal subcarriers for data transmission, and EDs are distinguished based on time segments (in each time segment, one ED occupies all subcarriers in a channel to transmit a complete data packet \cite{li2006orthogonal}), resulting in \textit{sequential and non-preemptive data transmission in the wireless network}. The computation resource capacity of a server is measured in computing units which are defined based on the resource type supported by the server, i.e., the computing unit of a CPU server is the CPU core, and that of a GPU server is the streaming multiprocessor. Each server $m_i \in \mathcal{M}^p$ comprises exactly one resource type (e.g., CPU, GPU, etc.) and provides a set of resource allocation options $\mathcal{C}_{i}$ for job processing, where $|\mathcal{C}_{i}| \le C$ for some integer $C$, and each option $c \in \mathcal{C}_{i}$ corresponds to some fraction of $\alpha_i$, i.e., $c \in (0, 1]$. Notably, servers with multiple resource types can be viewed as multiple co-located servers, each possessing a single resource type.
Given the potentially significant context switch overhead \cite{zhou2022online}, we consider \textit{non-preemptive job processing}. Besides, servers can run multiple jobs concurrently, provided their combined resource allocations do not exceed the server's capacity.



Let $\mathcal{N}$ be the set of $N$ jobs to be processed. Each job $n_j \in \mathcal{N}$ is represented by $\langle \theta_j^{in}$, $\theta_j^{out}, \gamma_j, \delta_j, \mathcal{M}^p_j, \psi_j \rangle$. Here, $\theta_j^{in}$ is the input data size, $\theta_j^{out}$ is the output data size, $\gamma_j$ is the release time, and $\delta_j$ is the absolute deadline of job $n_j$. Both $\theta_j^{in}$ and $\theta_j^{out}$ are measured in Megabyte (MB). Each job $n_j$ can be processed on a server with the required resource type; let $\mathcal{M}^p_j$ be the set of servers where $n_j$ can be processed. Due to the design of network rings, the accuracy requirement for trajectory estimation is reduced (i.e., we only need to determine if vehicles are under the coverage of network rings); thus, the ED trajectories can be estimated by mainstream navigation applications such as Google Maps \cite{fang2020constgat}. Based on the estimated trajectory of $n_j$'s ED, the accessible vAPs of $n_j$ can be defined by a set of at most $R$ \textit{ring coverage windows} $\psi_j$ throughout its lifetime $[\gamma_j,\delta_j]$. Each coverage window $T_{ir} \in \psi_j$ indicates that $n_j$'s ED is covered by network ring $m_{ir}$ during time window $T_{ir}$; the starting and ending time slots of $T_{ir}$, termed $st(T_{ir})$ and $et(T_{ir})$, depend on the estimated moving speed and direction of $n_j$'s ED. If $m_{ir}$ is chosen for offloading or downloading, we require that the operation must be completed within $T_{ir}$. We provide an example of the full cycle of an offloaded job in Fig. \ref{fig:system3}. Each job can either be processed locally or on servers. We use $d_j^{loc}$ to denote the local processing duration of job $n_j$, and $d_{jic}$ to denote the processing duration of job $n_j$ when it is processed on server $m_i$ with resource allocation option $c \in \mathcal{C}_i$. We assume that the ED of each job $n_j$ does not generate other jobs during $[\gamma_j,\delta_j]$ and
$\gamma_j + d_j^{loc} - 1 \leq \delta_j$ (ensuring local processing feasibility).   

\begin{table}
    \caption{Notation (Main Parameters and Variables)}
    \resizebox{\columnwidth}{!}{
    \begin{tabular}[t]{|m{0.04\textwidth}|p{0.4\textwidth}|}
        \hline
        Notation & Definition \\ \hline
        $\mathcal{M}^u$ & the set of $M^u$ vAPs with uplink channels \\ \hline
        $\mathcal{M}^p$ & the set of $M^p$ servers \\ \hline
        $\mathcal{M}^d$ & the set of $M^d$ vAPs with downlink channels \\ \hline
        $\mathcal{M}$ & $\mathcal{M} = \mathcal{M}^u \cup \mathcal{M}^p \cup \mathcal{M}^d$, set of machines (vAPs and servers); $m_i \in \mathcal{M}$ denotes a machine; $M = \max\{M^u, M^p, M^d\}$ \\ \hline
        $\alpha_i$ & the resource capacity of machine $m_i \in \mathcal{M}$ \\ \hline
        $\mathcal{C}_i$ & resource allocation options of server $m_i \in \mathcal{M}^p$; $|\mathcal{C}_i| \le C$ \\ \hline
        $m_{ir}$ & $r$-th network ring of vAP $m_i \in \mathcal{M}^u \cup \mathcal{M}^d$  \\ \hline
        $\mathcal{N}$ & the set of $N$ jobs, where $n_j \in \mathcal{N}$ denotes a job \\ \hline
        $\theta_j^{in}$ & input data size of job $n_j$ \\ \hline
        $\theta_j^{out}$ & output (result) data size of job $n_j$ \\ \hline 
        $\gamma_j$ & release time of job $n_j$ \\ \hline
        $\delta_j$ & absolute deadline of job $n_j$ \\ \hline
        $\Delta$ & $\Delta=\max_{n_j\in \mathcal{N}} \delta_j$, total number of time units for scheduling \\ \hline
        $\mathcal{M}^p_j$ & set of servers on which job $n_j$ can be processed; $\mathcal{M}^p_j \subseteq \mathcal{M}^p$ \\ \hline
        $\psi_j$ & $\{T_{ir}, ...\}$, set of ring coverage windows of job $n_j$; $|\psi_j| \le R$ \\ \hline
        $t^u_j$ & the offloading starting time of job $n_j$ \\ \hline
        $t^p_j$ & the processing starting time of job $n_j$ \\ \hline
        $t^d_j$ & the downloading starting time of job $n_j$ \\ \hline
        $d^u_j$ & offloading duration of job $n_j$ \\ \hline
        $d^p_j$ & processing duration of job $n_j$ \\ \hline
        $d^d_j$ & downloading duration of job $n_j$ \\ \hline
        $d^{u,p}_j$ & forwarding duration of job $n_j$'s input from its offloading vAP to its processing server \\ \hline
        $d^{p,d}_j$ & forwarding duration of job $n_j$'s output from its processing server to its downloading vAP \\ \hline
        $e_j^{save}$ & saved energy of job $n_j$’s ED by processing $n_j$ on a server \\ \hline
        $\ell$ &  $\ell$$\triangleq$$\langle m^u_\ell, m^p_\ell, m^d_\ell, I^u_\ell, I^p_\ell, I^d_\ell, c^p_\ell \rangle$ is a schedule instance. $m^u_\ell$, $m^p_\ell$ and $m^d_\ell$ are job mappings, $I^u_\ell$, $I^p_\ell$ and $I^d_\ell$ are operation intervals, and $c^p_\ell$ is allocated computation resource   \\ \hline
        $e(\ell)$ &  saved energy of schedule instance $\ell$  \\ \hline
        $\mathcal{L}$ & the set of all schedule instances of all jobs\\ \hline
        $\mathcal{L}_j$ & the set of schedule instances of $j$ \\ \hline
        $st(T)$ & start time of time interval/window $T$ \\ \hline
        $et(T)$ & end time of time interval/window $T$ \\ \hline
        \hline 
        $x(\ell)$ & binary selection variable of instance $\ell$ \\ \hline
    \end{tabular}
    }
    \label{table:notation}
\end{table}

Enabling jobs to be forwarded to different servers via the wired backhaul network after being offloaded offers advantages in balancing server workloads and mitigating wireless network coverage limitations. For instance, the servers co-located with the accessible vAPs of job $n_j$ may not possess the resource type required by $n_j$. We consider the \textit{backhaul network} connects vAPs and servers with optical cables and is enabled with Software Defined Network technology \cite{alameddine2019dynamic}. Such a backhaul network has enough bandwidth capacity \cite{xia2013commercial} to support data transmission with no communication contention and provides a fixed data transmission rate for each link. Then, given the topology of the backhaul network, the data forwarding duration between a vAP $m_i$ and a server $m_p$ can be defined as a function of the transmitted data size $\theta$. It comprises the data transmission duration over each link of the shortest path from vAP $m_i$ to server $m_p$; we denote this function as $\beta_{ip}(\theta)$. Note that $\beta_{ip}(\theta) = \beta_{pi}(\theta)$ and $\beta_{ii}(\theta) = 0$.


\subsection{Energy Consumption and Timing Model}\label{subsec:archiB}
\subsubsection{Local Processing} When job $n_j$ is processed locally, the processing duration is $d_{j}^{loc}$. Let $p_{j}^{loc}$ be the processing power. The energy consumption for processing job $n_j$ locally is then given by $e_{j}^{loc} = p_{j}^{loc} \cdot d_{j}^{loc}$.

\subsubsection{Remote Processing} Suppose job $n_j$ starts offloading to vAP $m_u$ at time $t^u_j$ during ring coverage window $T_{ur}$. Then, it is forwarded to server $m_p$ and starts processing at time $t^p_j$ with resource allocation $c \in \mathcal{C}_{p}$. After processing, job $n_j$ (result) is forwarded to vAP $m_d$ and starts downloading from $m_d$ to $n_j$'s ED at time $t^d_j$ during ring coverage window $T_{ds}$. Here, the \textit{subscript} $u,p$, and $d$ (e.g., $m_u$) are used as machine index, and the \textit{superscript} $u,p$, and $d$ (e.g., $t^u_j$) are used to denote offloading, processing, and downloading.

To ensure machine (vAP and server) capability, we have
\begin{equation}\label{eq:sys0}
    T_{ur} \in \psi_j,  T_{ds} \in \psi_j, \text{ and } m_p \in \mathcal{M}^p_j.
\end{equation}
Since job $n_j$ cannot start offloading before its release, 
\begin{equation}\label{eq:sys1}
    \gamma_j \le t^u_j.
\end{equation}
Based on Shannon's theorem \cite{shannon1984}, the data offloading rate is defined as $\eta_{j}^{u} = \alpha_{u} \cdot \log_2(1+p_{j}^{u} \cdot h_{ur}/{\sigma^2})$,
where $\alpha_u$ is the bandwidth capacity of vAP $m_u$, $p_j^u$ is the offloading power of ED, $h_{ur}$ is the channel gain in network ring $m_{ur}$, and $\sigma$ is the noise spectral density. The offloading duration $d_{j}^u$ and energy consumption $e_{j}^u$ are given by $ d_{j}^u = \theta^{in}_j / \eta_{j}^u \text{ and } e_{j}^u = p_{j}^u \cdot d_{j}^u.$ Furthermore, $t^u_j$ needs to satisfy
\begin{equation}\label{eq:sys2}
    st(T_{ur}) \le t^u_j \text{, and } t^u_j + d_{j}^u - 1 \le et(T_{ur}). 
\end{equation}

Since there is no bandwidth contention in the backhaul network, job $n_j$ can be immediately forwarded to server $m_p$ after being offloaded. Let $d_{j}^{u,p} = \beta_{up}(\theta^{in}_j)$ be job $n_j$'s forwarding duration from vAP $m_u$ to server $m_p$. As job $n_j$ cannot start processing before it arrives $m_p$, $t^p_j$ needs to satisfy
\begin{equation}\label{eq:sys3}
    t^u_j + d_{j}^u + d_{j}^{u,p} \le t^p_j.
\end{equation}
Let $d_j^p = d_{jpc}$ denote the processing duration of job $n_j$ on server $m_p$. Once job $n_j$ is processed, the result can be immediately forwarded to its downloading vAP $m_d$. Let $d_j^{p,d} = \beta_{pd}(\theta^{out}_j)$ denote this forwarding duration from server $m_p$ to vAP $m_d$. Job $n_j$ cannot start downloading from vAP $m_d$ before it arrives at $m_d$, and hence $t^d_j$ needs to satisfy
\begin{equation}\label{eq:sys4}
    t^p_j + d_j^p + d_j^{p,d} \le t^d_j.
\end{equation}
Each ED requires a minimum power to ensure the necessary sensitivity to \textit{receive} the wireless signal sent from a vAP \cite{wang2006realistic}. Let $p_j^d$ be the downloading power of job $n_j$'s ED. Let $\eta_j^d = \alpha_{d} \cdot \log_2(1+ \textit{SNR}_{d})$ be the data downloading rate of $n_j$, where $\textit{SNR}_{d}$ is the signal-to-noise ratio and depends on vAP $m_d$ since it is the data transmitter. 
Thus, the downloading duration $d_{j}^d$ and energy consumption $e_{j}^d$ for $n_j$ are given by $d_{j}^d = \theta^{out}_j / \eta_{j}^d \text{ and } e_{j}^d = p_{j}^d \cdot d_{j}^d$. Since $n_j$ can only be downloaded during ring coverage window $T_{ds}$, $t^d_j$ must satisfy
\begin{equation}\label{eq:sys5}
    st(T_{ds}) \le t^d_j \text{, and } t^d_j + d_j^d - 1 \le et(T_{ds}). 
\end{equation}
To meet the job deadline, $t^d_j$ also needs to satisfy
\begin{equation}\label{eq:sys6}
    t^d_j + d_j^d - 1 \le \delta_j.
\end{equation}
We omit the vAP handover delay for switching from the offloading vAP to the downloading vAP (if they are different), as this delay is typically shorter than the job processing time (i.e., the vAP handover can be completed before the job processing finishes).
The saved energy for job $n_j$'s ED by processing job $n_j$ on server $m_p$ is defined as $e^{save}_j = e_j^{loc} - e^u_j -  e^d_j$.

\subsection{Problem Formulation}\label{subsec:formulation}
\textbf{Problem Definition}: Let $\Delta=\max_{n_j\in \mathcal{N}} \delta_j$. This paper aims to find a schedule that satisfies machine capability constraints, resource capacity constraints, and timing constraints for all jobs over the $\Delta$ time units to maximize the total saved energy for EDs; we refer to this problem as the Energy Maximization Job Scheduling Problem ($\mathsf{EMJS}$). To solve $\mathsf{EMJS}$, we need to determine the job mapping (where each job is offloaded, processed and downloaded), computation resource allocation on server, and the starting times for each \textit{operation} (offloading, processing, and downloading). 
The \textit{machine capability constraint} is specified by Eqs. \eqref{eq:sys0}, \eqref{eq:sys2} and \eqref{eq:sys5}. The \textit{resource capacity constraint} ensures that at any given time instant, the total resource allocated to all the jobs by any machine does not exceed its capacity.
The \textit{timing constraints} are specified by Eqs. \eqref{eq:sys1}, \eqref{eq:sys3}, \eqref{eq:sys4}, and \eqref{eq:sys6}. 

\textbf{Problem Complexity}: When job executions are sequential on servers (no fractional resource allocation) and the MEC comprises one offloading vAP, one server and one downloading vAP, the resulting problem is a \emph{three-machine flow shop problem}, known to be NP-Hard and challenging to be approximated within $\mathcal{O}(3/ \log 3)$ unless P = NP \cite{ben2020maximizing}. Since $\mathsf{EMJS}$ considers a more general MEC and multiple computation resource allocation options, it is at least as hard as the three-machine flow shop problem.

\begin{definition}[Schedule Instance]\label{def:instance}
    A schedule instance of job $n_j$ is defined as $\ell \triangleq \langle m^u_\ell, m^p_\ell, m^d_\ell, I^u_\ell, I^p_\ell, I^d_\ell, c^p_\ell \rangle$, which represents the scenario where job $n_j$ offloads to vAP $m^u_\ell$ in the operation interval $I^u_\ell$ ($I^u_\ell \triangleq [t^u_j, t^u_j + d_j^u-1]$), executes on server $m^p_\ell$ with computation resource allocation $c^p_\ell \in \mathcal{C}_{\ell}$ in the operation interval $I^p_\ell$ ($I^p_\ell \triangleq [t^p_j, t^p_j + d_j^p-1]$), and downloads its result from vAP $m^d_\ell$ in the operation interval $I^d_\ell$ ($I^d_\ell \triangleq [t^d_j, t^d_j + d_j^d-1]$). Moreover, $\ell$ needs to satisfy 
    \begin{enumerate}[(i)]
        \item machine capability constraint: Eqs. \eqref{eq:sys0}, \eqref{eq:sys2}, and \eqref{eq:sys5}. 
        \item timing constraint: Eqs. \eqref{eq:sys1}, \eqref{eq:sys3}, \eqref{eq:sys4}, and \eqref{eq:sys6}.
    \end{enumerate}
    For ease of presentation, we use notation $I^*_\ell$ (likewise $m^*_\ell$) to generically denote any of the three operation intervals (likewise machines) of $\ell$.
\end{definition}

Let $\ell$ be a schedule instance of job $n_j$. Let $e(\ell) = e_j^{save}$ be the \textit{saved energy} for $n_j$'s ED when $n_j$ is scheduled following $\ell$. For any set of schedule instances $\mathcal{S}$, let $e(\mathcal{S}) = \sum_{\ell \in \mathcal{S}} e(\ell)$. For any machine $m_i$ and time instant $t$, let $\mathds{1}_{i}(t,I^*_\ell)$ be an indicator function, where $\mathds{1}_{i}(t,I^*_\ell) = 1$ if and only if the operation interval $I^*_\ell$ is active at time $t$ ($t \in I^*_\ell$) and $m_i = m^*_\ell$.

In offline $\mathsf{EMJS}$, information about all the jobs is available apriori. Thus, we can enumerate all possible schedule instances for each job $n_j \in \mathcal{N}$. Let $\mathcal{L}$ be the set of all possible schedule instances for all the jobs in $\mathcal{N}$, and $\mathcal{L}_j$ be the set of all possible schedule instances for any job $n_j$. Note that we are only interested in those schedule instances $\ell$ with positive $e(\ell)$; thus, $\mathcal{L}$ and $\mathcal{L}_j$ contain only those schedule instances with positive saved energy. Since each job has at most $R$ ring coverage windows and each server has at most $C$ resource allocation options, we have at most $NMCR^2\Delta^3$ schedule instances in $\mathcal{L}$. Because $\Delta$ depends on the value of $\delta_j$, the number of schedule instances in $\mathcal{L}$ is pseudo-polynomial to the input size.
For each schedule instance $\ell \in \mathcal{L}$, let $x_\ell \in \{0,1\}$ be the \textit{selection variable} of $\ell$, where $x_\ell = 1$ if and only if $\ell$ is selected in the solution. Then, we formulate offline $\mathsf{EMJS}$ as follows.
\begin{equation}\label{eq:emjs}
    (\mathsf{EMJS} \mbox{ Offline}) \ \ \max { \sum_{\ell \in \mathcal{L}} e(\ell) \cdot x_\ell}
\end{equation}
subject to:
\addtocounter{equation}{-1}
\begin{subequations}
    \allowdisplaybreaks
    \begin{align}
        \sum_{\substack{\ell \in \mathcal{L}, I^*_\ell \in \ell,\\ \mathds{1}_{i}(t,I^*_\ell)=1}} \scaleto{x_\ell \le 1,  \forall 1 \le t \le \Delta,  \forall m_i \in \mathcal{M}^u \cup \mathcal{M}^d}{10pt}\label{eq:emjs1} \\
        { \sum_{\substack{\ell \in \mathcal{L},  \mathds{1}_{i}(t,I^p_\ell)=1}} x_\ell \cdot c^p_\ell \le 1}, \forall 1 \le t \le \Delta,  \forall m_i \in \mathcal{M}^p \label{eq:emjs2} \\
        { \sum_{\ell \in \mathcal{L}_j} x_\ell \le 1}, \ \ \forall n_j \in \mathcal{N} \label{eq:emjs3} \\
        x_\ell \in \{0,1\}, \ \ \forall \ell \in \mathcal{L} \label{eq:emjs4}
    \end{align}
\end{subequations}
Eqs. \eqref{eq:emjs1} and \eqref{eq:emjs2} are the resource capacity constraints for vAPs and servers, respectively. Constraint \eqref{eq:emjs3} guarantees that at most one schedule instance of each job is selected in the solution. Note that the machine capability and timing constraints have already been considered in the definition of schedule instances. The formulation contains $NMCR^2\Delta^3$ variables and ($3M\Delta+J$) constraints, which is a pseudo-polynomial of the input size.

\section{An Approximation Algorithm for Offline EMJS}\label{sec:approximation}
This section presents an approximation algorithm for offline $\mathsf{EMJS}$, called the Light-Heavy Job Scheduling Algorithm ($\mathtt{LHJS}$). (For ease of presentation, whenever we mention $\mathsf{EMJS}$ in this section it always refers to the offline version.) 
For each schedule instance $\ell \triangleq \langle m^u_\ell, m^p_\ell, m^d_\ell, I^u_\ell, I^p_\ell, I^d_\ell, c^p_\ell \rangle$ of job $n_j$, $0 < c^p_\ell \le 1$ denotes the computation resource allocation on server $m^p_\ell$. We refer to $\ell$ as a \textit{light schedule instance} if $c^p_\ell \le \frac{1}{2}$, and as a \textit{heavy schedule instance} otherwise. Further, for ease of presentation, 
 we use $c^*_\ell$ to generically denote the resource allocation corresponding to any operation interval $I^*_\ell$ in $\ell$; if $I^*_\ell = I^p_\ell$ then $c^*_\ell = c^p_\ell$, and if $I^*_\ell = I^u_\ell$ or $I^d_\ell$ then $c^*_\ell = 1$ denoting the full bandwidth allocation for offload/download. Finally, for a set of operation intervals $\mathcal{I}$, let $c(\mathcal{I}) = \sum_{I^*_\ell \in \mathcal{I}} c^*_\ell$.

In $\mathtt{LHJS}$ (Algorithm \ref{alg:lhjs}), we first divide $\mathcal{L}$ into two sets: the set of all light schedule instances $\mathcal{L}_{L}$ and the set of all heavy schedule instances $\mathcal{L}_{H}$. Next, we apply $\mathtt{RandRound}$ (Section \ref{sec:RandRounding}) to obtain a solution for $\mathcal{L}_{L}$ and use $\mathtt{SortSched}$ (Section \ref{sec:SortSched}) to obtain a solution for $\mathcal{L}_{H}$. Finally, we select the solution with a higher energy saving as the final solution for $\mathsf{EMJS}$. Here, we use vectors $\mathbf{y}$ and $\mathbf{z}$ to represent the optimal fractional solutions of LP problems $\mathsf{LIS}$ and $\mathsf{HIS}$, respectively, to differentiate from the general solution $\mathbf{x}$. Notably, solving an LP problem with $n$ variables optimally takes time $\mathcal{O}(n^3)$ \cite{vaidya1987an}, and is much faster than solving an ILP problem, which takes exponential time.
Later in this section, we show that $\mathtt{LHJS}$ achieves a constant approximation ratio for $\mathsf{EMJS}$.

\textit{Motivation for $\mathtt{LHJS}$.} Simultaneously considering schedule instances with computation resource allocations spanning $0$ to $1$ is very challenging to approximate. By scheduling $\mathcal{L}_{L}$ and $\mathcal{L}_{H}$ separately, we can obtain two additional properties that aid in deriving an approximation ratio for $\mathsf{EMJS}$:
\begin{enumerate}[(i)]
    \item When scheduling $\mathcal{L}_{L}$, if $\ell \in \mathcal{L}_{L}$ cannot be included in the solution due to conflict in $I^p_\ell$, then at least $\frac{1}{2}$ of the resource of $m^p_\ell$ has been allocated to other schedule instances in the solution at some time $t \in I^p_\ell$.
    \item When scheduling $\mathcal{L}_{H}$, since $c^p_\ell > \frac{1}{2}$ for all $\ell \in \mathcal{L}_{H}$, jobs cannot be scheduled in parallel on any server in the solution, i.e., two intervals $I^p_{\ell_1}$ and $I^p_{\ell_2}$ in the solution cannot overlap if $m^p_{\ell_1} = m^p_{\ell_2}$.
\end{enumerate}

\begin{algorithm}[tb]
    \caption{Light-Heavy Job Scheduling ($\mathtt{LHJS}$)}\label{alg:lhjs}
    \hspace*{\algorithmicindent} \textbf{Input}: $\mathcal{N}$, $ \mathcal{M}$  \\
    \hspace*{\algorithmicindent} \textbf{Output}: $\mathcal{S}$
    \begin{algorithmic}[1]
        \STATE Define $\scaleto{\mathcal{L}}{6.5pt}$ by enumerating all possible schedule instances;
        \STATE Divide $\mathcal{L}$ into $\mathcal{L}_{L}$ and $\mathcal{L}_{H}$;
        \STATE Define an LP formulation ($\mathsf{LIS}$) for $\mathcal{L}_{L}$ based on $\mathsf{EMJS}$; let $\mathbf{y}$ be an optimal fractional solution of $\mathsf{LIS}$.
        \STATE $\mathcal{S}_{L} \leftarrow \mathtt{RandRound}(\mathbf{y})$;
        \STATE Define an LP formulation ($\mathsf{HIS}$) for $\mathcal{L}_{H}$ based on $\mathsf{EMJS}$; let $\mathbf{z}$ be an optimal fractional solution of $\mathsf{HIS}$;
        \STATE $\mathcal{S}_{H} \leftarrow \mathtt{SortSched}(\mathbf{z})$;
        \STATE \textbf{if} $e(\mathcal{S}_{L}) \ge e(\mathcal{S}_{H})$ \textbf{do} $\mathcal{S} \leftarrow \mathcal{S}_{L}$;\\ \textbf{else} $\mathcal{S} \leftarrow \mathcal{S}_{H}$.
    \end{algorithmic}
\end{algorithm}

\subsection{The RandRound Algorithm for Light Schedule Instances} \label{sec:RandRounding}
\begin{figure}[tb]
    \centering
    \includegraphics[width=0.75\linewidth, page=1]{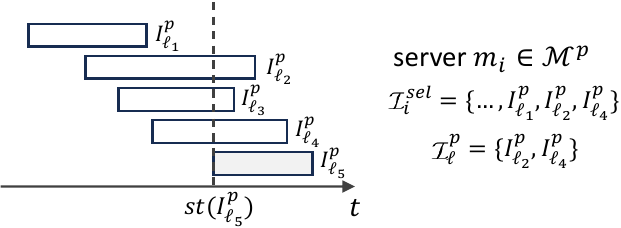}
    \caption{An example of operation interval selection for server $m_i \in \mathcal{M}^p$, where we add $I_{\ell_5}^p$ into $\mathcal{I}_i^{sel}$ if $c^p_\ell + c(\mathcal{I}_\ell^p) \le 1$.}
    \label{fig:randRound}
\end{figure}

This subsection presents a randomized rounding algorithm, called $\mathtt{RandRound}$, to obtain a feasible scheduling solution for $\mathcal{L}_{L}$. 
Based on the formulation of $\mathsf{EMJS}$, we define a relaxed LP formulation corresponding to $\mathcal{L}_{L}$, called $\mathsf{LIS}$, by relaxing $x_\ell$ into a continuous variable in the range $[0,1]$. 
\begin{equation}\label{eq:ljs}
    (\mathsf{LIS})\ \ \max {\textstyle \sum_{\ell \in \mathcal{L}_{L}} e(\ell) \cdot x_\ell}
\end{equation}
subject to Eqs. \eqref{eq:emjs1}, \eqref{eq:emjs2}, \eqref{eq:emjs3}, and  $x(\ell) \ge 0, \ \ \forall \ell \in \mathcal{L}_{L}$. 
We denote the optimal fractional solution of $\mathsf{LIS}$ as $\mathbf{y}$. Let $y(\mathcal{L}_j) = \sum_{\ell \in \mathcal{L}_j} y_\ell$. Next, we use $\mathtt{RandRound}$ (Algorithm \ref{alg:rr}) to obtain a feasible scheduling solution for $\mathsf{EMJS}$ by rounding $\mathbf{y}$. Specifically, the function $Random(0,1)$ in lines $3$ and $7$ of $\mathtt{RandRound}$ samples a value of range $[0,1]$ \textit{independently at random}, and the value of $\kappa$ in line $4$ of $\mathtt{RandRound}$ is determined in the proof of Theorem \ref{theorem:ljs1}. Furthermore, we provide an example for operation interval selection (lines $11$--$16$) in Fig. \ref{fig:randRound}.
In the following, we show that $\mathcal{S}_{L}$ obtained by $\mathtt{RandRound}$ is a feasible solution to $\mathsf{EMJS}$, and derive the approximation ratio for $\mathtt{RandRound}$.

\begin{algorithm}[tb]
    \caption{Randomized Rounding ($\mathtt{RandRound}$)}\label{alg:rr}
    \hspace*{\algorithmicindent} \textbf{Input}: $\mathbf{y}$ (an optimal fractional solution of $\mathsf{LIS}$) \\
    \hspace*{\algorithmicindent} \textbf{Output}: $\mathcal{S}_{L}$
    \begin{algorithmic}[1]    
    \STATE $\mathcal{N}^{sel} \leftarrow \emptyset, \mathcal{L}^{sel} \leftarrow \emptyset, \mathcal{I}^{sel} \leftarrow \emptyset, \mathcal{S}_{L} \leftarrow \emptyset$;  
    
    {\small $/\ast$ \textit{Step 1: select job $n_j$ with probability $y(\mathcal{L}_j)/\kappa$} $\ast/$}
    \FOR{$n_j \in \mathcal{N}$}
    \STATE $val_1 \leftarrow$ $Random(0,1)$;
    \STATE \textbf{if} {$val_1 \le {y(\mathcal{L}_j)}/{\kappa}$} \textbf{do} $\mathcal{N}^{sel} \leftarrow \mathcal{N}^{sel} \cup \{n_j\}$;
    \ENDFOR

    {\small $/\ast$ \textit{Step 2: select one $\ell$ with prob. $y_\ell/y(\mathcal{L}_j)$ for $n_j \in \mathcal{N}^{sel}$} $\ast/$}
    \FOR{$n_j \in \mathcal{N}^{sel}$}
    \STATE \textbf{for all} $\ell \in \mathcal{L}_j$ \textbf{do} $\tilde{y}_\ell \leftarrow y_\ell/y(\mathcal{L}_j)$;
    \STATE $val_2 \leftarrow$ $Random(0,1)$;
    
    {\small $/\ast$ \textit{let $\ell^{k}$ be the $k$-th instance in $\mathcal{L}_j$} $\ast/$} 
    \IF{$\sum_{k=1}^{s-1} \tilde{y}_{\ell^{k}} < val_2 \le \sum_{k=1}^{s} \tilde{y}_{\ell^{k}}$}
    \STATE $\mathcal{L}^{sel} \leftarrow \mathcal{L}^{sel} \cup \{\ell^{s}\}$;   
    \ENDIF
    \ENDFOR

    {\small $/\ast$ \textit{Step 3: select operation intervals} $\ast/$}
    \FOR{$m_i \in \mathcal{M}$}
    \STATE $\mathcal{I}^{sel}_i \leftarrow \emptyset$; 
    \STATE $\mathcal{I}_i \leftarrow \{I^*_\ell \mid \ell \in \mathcal{L}^{sel} \text{ and } \exists t, \mathds{1}_{i}(t, I^*_\ell)=1\}$;
    \STATE Sort all $ I^*_\ell \in \mathcal{I}_i$ in ascending order of $st(I^*_\ell)$;
    \FOR{$ I^*_\ell \in \mathcal{I}_i$ {\small \texttt{(from left to right)}}}
    \STATE $\mathcal{I}^*_\ell \leftarrow \{I^*_{\ell'} \in \mathcal{I}^{sel}_i \mid \mathds{1}_{m}(st(I^*_\ell), I^*_{\ell'})=1 \}$;
    \STATE \textbf{if} $c^*_\ell + c(\mathcal{I}^*_\ell) \le 1$ \textbf{do} $\mathcal{I}^{sel}_i \leftarrow \mathcal{I}^{sel}_i \cup \{I^*_\ell\}$;
    \ENDFOR
    \STATE $\mathcal{I}^{sel} \leftarrow \mathcal{I}^{sel} \cup \mathcal{I}^{sel}_i$;
    \ENDFOR

    \STATE $\mathcal{S}_{L} \leftarrow \{\ell \in  \mathcal{L}^{sel} \mid \forall I^*_\ell \in \ell, I^*_\ell \in \mathcal{I}^{sel} \}$.
    \end{algorithmic}
\end{algorithm}

\begin{lemma}\label{lemma:ljs1}
    The output $\mathcal{S}_{L}$ of $\mathtt{RandRound}$ is a feasible scheduling solution for $\mathsf{EMJS}$.
\end{lemma}

\begin{IEEEproof}
    In lines $5$--$9$, we choose at most one schedule instance for each job, so $\mathcal{S}_{L}$ satisfies constraint \eqref{eq:emjs2} of $\mathsf{EMJS}$. In line $16$, we only add an operation interval into set $\mathcal{I}^{sel}$ when the resource capacity constraint is not violated. Thus, $\mathcal{S}_{L}$ also satisfies the constraints \eqref{eq:emjs1} and \eqref{eq:emjs2} of $\mathsf{EMJS}$.
\end{IEEEproof}

\begin{theorem}\label{theorem:ljs1}
    Suppose $\mathbf{y}$ is an optimal solution to $\mathsf{LIS}$, and $\mathcal{S}_{L}$ is the schedule generated by $\mathtt{RandRound}$.
    Let $\kappa = 8$ (line $4$ of $\mathtt{RandRound}$) and $e(\mathbf{y}) = \sum_{\ell \in \mathcal{L}_{L}} e(\ell) \cdot y_\ell$. Then, the expected value of $e(\mathcal{S}_{L})$ satisfies the condition $\mathbf{E}[e(\mathcal{S}_{L})] \ge e(\mathbf{y})/16$.
\end{theorem}

\begin{IEEEproof}
    Let $\ell \triangleq \langle m^u_\ell, m^p_\ell, m^d_\ell, I^u_\ell, I^p_\ell, I^d_\ell, c^p_\ell \rangle$ be a light schedule instance of job $n_j$. The probability of $\ell \in \mathcal{S}_{L}$,
    \begin{equation}\label{eq:ljs1}
        \text{Pr}[\ell \in \mathcal{S}_{L}] = \text{Pr}[\ell \in \mathcal{L}^{sel}] \cdot \text{Pr}[\ell \in \mathcal{S}_{L} \mid \ell \in \mathcal{L}^{sel}].
    \end{equation}
    Based on lines $3$--$9$, we have $\text{Pr}[n_j \in \mathcal{N}^{sel}] = {y(\mathcal{L}_j)}/{\kappa}$ and $\text{Pr}[\ell \in \mathcal{L}^{sel} \mid n_j \in \mathcal{N}^{sel}] = {y_\ell}/{y(\mathcal{L}_j)}$.
    Thus, the first term on the right-hand side (RHS) of Eq. \eqref{eq:ljs1},
    \begin{equation}\label{eq:ljs2}
        \scaleto{\text{Pr}[\ell \in \mathcal{L}^{sel}] = \text{Pr}[n_j \in \mathcal{N}^{sel}] \cdot \text{Pr}[\ell \in \mathcal{L}^{sel} \mid n_j \in \mathcal{N}^{sel}] }{10.4pt} = { \frac{y_\ell}{\kappa}}.
    \end{equation}
    For the second term on the RHS of Eq. \eqref{eq:ljs1},
    \begin{equation}\label{eq:ljs3}
        \begin{aligned}
            \scaleto{\text{Pr}[\ell \in \mathcal{S}_{L}}{9.5pt} \mid & \scaleto{\ell \in  \mathcal{L}^{sel}]  = \text{Pr}[\forall I^*_\ell \in \ell, I^*_\ell \in \mathcal{I}^{sel} \mid \ell \in \mathcal{L}^{sel}]}{11pt} \\
             &= 1- \text{Pr}[\exists I^*_\ell \in \ell, I^*_\ell \notin \mathcal{I}^{sel} \mid \ell \in \mathcal{L}^{sel}] \\
            & \ge {\textstyle 1- \sum_{I^*_\ell \in \ell} \text{Pr}[I^*_\ell \notin \mathcal{I}^{sel} \mid \ell \in \mathcal{L}^{sel}]. }
        \end{aligned}
    \end{equation}
    The inequality in Eq. \eqref{eq:ljs3} follows from the union bound. 
    Then, we derive a bound on each term of the summation in Eq. \eqref{eq:ljs3}.
    
    We first analyze $\text{Pr}[I^u_\ell \notin \mathcal{I}^{sel} \mid \ell \in \mathcal{L}^{sel}]$. When $I^*_\ell = I^u_\ell$, $c^*_\ell = 1$ and the offloading interval cannot overlap with each other at any vAP in a feasible scheduling solution. Thus,
    \begin{subequations}
    \begin{align}
        &\text{Pr}[I^u_\ell \notin \mathcal{I}^{sel} \mid \ell \in \mathcal{L}^{sel}]  \\
        =  &\text{Pr}[\exists I^u_{\ell'} \in \mathcal{I}^u_{\ell}, I^u_{\ell'} \in \mathcal{I}^{sel} \mid \ell \in \mathcal{L}^{sel}] && \text{(lines $15$--$16$)}\label{eq:ljs4a}  \\
        \le & {\textstyle \sum_{I^u_{\ell'} \in \mathcal{I}^u_\ell} \text{Pr}[ I^u_{\ell'} \in \mathcal{I}^{sel} \mid \ell \in \mathcal{L}^{sel}]} && \text{(union bound)}\label{eq:ljs4b}  \\
        \le  & {\textstyle \sum_{I^u_{\ell'} \in \mathcal{I}^u_\ell} \text{Pr}[\ell' \in \mathcal{L}^{sel} \mid \ell \in \mathcal{L}^{sel}]} && \text{(line $12$)} \label{eq:ljs4c}  \\
        =  &{\textstyle \sum_{I^u_{\ell'} \in \mathcal{I}^u_\ell} \text{Pr}[\ell' \in \mathcal{L}^{sel}]} && \text{(line $3$)} \label{eq:ljs4d}  \\
        =  &{\textstyle \sum_{I^u_{\ell'} \in \mathcal{I}^u_\ell} \frac{y_{\ell'}}{\kappa} \le  \frac{1}{\kappa}} && \text{(Eqs. \eqref{eq:ljs2})} \label{eq:ljs4e} 
    \end{align}
    \end{subequations}
    The last inequality in Eq. \eqref{eq:ljs4e} holds due to Eq.\eqref{eq:emjs1} and the fact that all $I^u_{\ell'} \in \mathcal{I}^u_\ell$ are active at time $st(I^u_\ell)$.
    Following a similar derivation, we have $\text{Pr}[I^d_\ell \notin \mathcal{I}^{sel} \mid \ell \in \mathcal{L}^{sel}] \le \frac{1}{\kappa}$.

    Next, we analyze $\text{Pr}[I^p_\ell \notin \mathcal{I}^{sel} \mid \ell \in \mathcal{L}^{sel}]$. Since $\ell \in \mathcal{L}_{L}$, $0< c^p_\ell \le \frac{1}{2}$. If $I^p_\ell$ cannot be added to $\mathcal{L}^{sel}$, we have $c(\mathcal{I}^p_\ell) \ge \frac{1}{2}$ at the time $st(I^p_\ell)$. Hence,
    \begin{equation}\label{eq:ljs5}
        {\textstyle \text{Pr}[I^p_\ell \notin \mathcal{I}^{sel} \mid \ell \in \mathcal{L}^{sel}] \le \text{Pr}[c(\mathcal{I}^p_\ell) \ge \frac{1}{2} \mid  \ell \in \mathcal{L}^{sel}].}
    \end{equation}
    Next, we derive a bound on the RHS of Eq. \eqref{eq:ljs5}. The expectation of $c(\mathcal{I}^p_\ell)$, 
    \[{\textstyle \mathbf{E}[c(\mathcal{I}^p_\ell) \mid \ell \in \mathcal{L}^{sel}] = \sum_{I^p_{\ell'} \in \mathcal{I}^p_\ell} c^p_{\ell'} \cdot \text{Pr}[I^p_{\ell'} \in \mathcal{I}^{sel} \mid \ell \in \mathcal{L}^{sel}].}\]
    Following a similar induction as Eqs.\eqref{eq:ljs4c} $\sim$ \eqref{eq:ljs4e} (the last inequality in Eq. \eqref{eq:ljs4e} will then hold due to Eq.\eqref{eq:emjs2}), we have $\mathbf{E}[c(\mathcal{I}^p_\ell) \mid \ell \in \mathcal{L}^{sel}] \le \frac{1}{\kappa}$.
    Then, by applying Markov's inequality (i.e., $\text{Pr}[X \ge a] \le {\mathbf{E}[X]}/{a}$), we have 
    \[{\textstyle \text{Pr}[c(\mathcal{I}^p_\ell) \ge \frac{1}{2} \mid \ell \in \mathcal{L}^{sel}] \le \frac{1}{\kappa} \div \frac{1}{2} = \frac{2}{\kappa}.}\] 
    Thus, by Eq. \eqref{eq:ljs5}, we have $\text{Pr}[I^p_\ell \notin \mathcal{I}^{sel} \mid \ell \in \mathcal{L}^{sel}] \le {2}/{\kappa}$. 
    
    Substituting the results for $I^u_\ell, I^p_\ell$, and $I^d_\ell$ in Eq. \eqref{eq:ljs3}, we have $\text{Pr}[\ell \in \mathcal{S}_{L} \mid \ell \in \mathcal{L}^{sel}] \ge 1- {4}/{\kappa}$.
    Based on Eq. \eqref{eq:ljs1}, $\text{Pr}[\ell \in \mathcal{S}_{L}] \ge y_\ell \cdot (\frac{1}{\kappa} - \frac{4}{\kappa^2})$. When $\kappa = 8$, $(\frac{1}{\kappa} - \frac{4}{\kappa^2})$ reaches its maximum ${1}/{16}$. Thus, the expected value of $e(\mathcal{S}_{L}), \mathbf{E}[e(\mathcal{S}_{L})]$
    $=\sum_{\ell \in \mathcal{L}_{L}} \text{Pr}[\ell \in \mathcal{S}_{L}] e(\ell) \ge \frac{1}{16} \sum_{\ell \in \mathcal{L}_{L}}  e(\ell) y_\ell = {e(\mathbf{y})}/{16}.$ 
\end{IEEEproof}

\subsection{The SortSched Algorithm for the Heavy schedule instances} \label{sec:SortSched}
In this subsection, we present an algorithm based on partial elimination ordering and a fractional local ratio method, referred to as $\mathtt{SortSched}$, to obtain a feasible scheduling solution for $\mathcal{L}_{H}$. Based on the formulation of $\mathsf{EMJS}$, we first define a relaxed LP formulation corresponding to $\mathcal{L}_{H}$, called $\mathsf{HIS}$. When only considering $\mathcal{L}_{H}$, no two operation intervals related to the same machine $m \in \mathcal{M}$ can overlap at any time. Thus, $\mathsf{HIS}$ can be defined as follows.
\begin{equation}\label{eq:hjs}
    (\mathsf{HIS}) \ \ \max {\textstyle \sum_{\ell \in \mathcal{L}_{H}} e(\ell) \cdot x_\ell} \text{ \ \ \ \ subject to: }
\end{equation}
\begin{equation}\label{eq:hjsa}
    {\textstyle \sum_{\substack{\ell \in \mathcal{L}_{H}, I^*_\ell \in \ell,  \mathds{1}_{i}(t,I^*_\ell)=1}} x_\ell \le 1, \forall 1 \le t \le \Delta, \forall m_i \in \mathcal{M}},
\end{equation}
Eq. \eqref{eq:emjs3}, and $x(\ell) \ge 0, \forall \ell \in \mathcal{L}_{H}$. For a heavy schedule instance $\ell$ of job $n_j$, let $\mathcal{L}(\ell)$ be the set of heavy schedule instances belonging to the same job $n_j$, and $\mathcal{A}(\ell)$ be the set of heavy schedule instances whose operation intervals (at least one) overlap with $\ell$ but do not belong to $\mathcal{L}(\ell)$. We first show the following property for any feasible solution of $\mathsf{HIS}$.

\begin{proposition}\label{prop:hjs1}
    Let $\mathbf{x}$ be any feasible solution to $\mathsf{HIS}$. There is a schedule instance $\ell$ satisfying $x_\ell + \sum_{\ell' \in \mathcal{A}(\ell)}x_{\ell'} \le 6.$
\end{proposition}
\begin{IEEEproof} 
    For two intersecting schedule instances $\ell$ and $\ell'$ (i.e., $\ell' \in \mathcal{A}(\ell)$ and $\ell \in \mathcal{A}(\ell')$), let $q(\ell, \ell') = x_\ell\cdot x_{\ell'}$. Besides, $q(\ell, \ell) = (x_\ell)^2$. For an operation interval $I^*_\ell \in \ell$, let $\mathcal{R}(I^*_\ell)$ be the set of schedule instances in $\mathcal{L}_H$ that have an operation interval intersecting the right end of $I^*_\ell$ (including $\ell$ itself). Consider $\sum_{\ell \in \mathcal{L}_H}(q(\ell, \ell) + \sum_{\ell' \in \mathcal{A}(\ell)}q(\ell, \ell'))$. We first obtain an upper bound of this sum as follows.
    
    For each operation interval $I^*_\ell$ of $\ell$, we sum up $q(\ell, \ell')$ for all schedule instances $\ell'$ having at least one operation interval that intersects with $I^*_\ell$ (including $\ell$ itself). We argue that \textit{it suffices to sum up $q(\ell, \ell')$ only for schedule instances $\ell' \in \mathcal{R}(I^*_\ell)$ and then multiply the total sum by $2$}. This is because if an operation interval $I^*_{\ell'}$ of schedule instance $\ell'$ intersects with $I^*_\ell$, we have either $\ell' \in \mathcal{R}(I^*_\ell)$ or $\ell \in \mathcal{R}(I^*_{\ell'})$. Since $q(\ell, \ell') = q(\ell', \ell)$,
    \begin{equation}\label{eq:hjs2}
        \scaleto{\sum_{\ell \in \mathcal{L}_H}(q(\ell, \ell) + \sum_{\ell' \in \mathcal{A}(\ell)}q(\ell, \ell')) \le 2\sum_{\ell \in \mathcal{L}_H}\sum_{I^*_\ell \in \ell}\sum_{\ell' \in \mathcal{R}(I^*_\ell)}q(\ell, \ell').}{23.5pt}
    \end{equation}
    Based on constraint \eqref{eq:hjsa} of $\mathsf{HIS}$ and the definition of $q(\ell, \ell')$,
    \begin{equation}\label{eq:hjs3}
        {\textstyle \sum_{\ell' \in \mathcal{R}(I^*_\ell)}q(\ell, \ell') \le x_\ell \cdot \sum_{\ell' \in \mathcal{R}(I^*_\ell)}x_{\ell'} \le x_\ell.}
    \end{equation}
    Using Eqs. \eqref{eq:hjs2} and \eqref{eq:hjs3}, and the fact that each schedule instance has $3$ operation intervals, we can get
    \begin{equation}\label{eq:hjs4}
        \scaleto{\sum_{\ell \in \mathcal{L}_H}(q(\ell, \ell) + \sum_{\ell' \in \mathcal{A}(\ell)} q(\ell, \ell')) \le 2\sum_{\ell \in \mathcal{L}_H}\sum_{I^*_\ell \in \ell} x_\ell \le 6 \sum_{\ell \in \mathcal{L}_H}x_\ell.}{23.5pt}
    \end{equation}
    Therefore, we can conclude that there exists at least one (otherwise Eq. \eqref{eq:hjs4} will not hold) schedule instance $\ell$ satisfies
    \begin{equation}\label{eq:hjs5}
        q(\ell, \ell) + \sum_{\ell' \in \mathcal{A}(\ell)}q(\ell, \ell') = (x_\ell)^2 + \sum_{\ell' \in \mathcal{A}(\ell)} x_{\ell'} \cdot x_\ell \le 6 \cdot x_\ell.
    \end{equation}
    This lemma is proved by factoring out $x_\ell$ from both sides of the inequality of Eq. \eqref{eq:hjs5}.
\end{IEEEproof}

Let $\mathbf{z}$ be an optimal fractional solution to the LP problem $\mathsf{HIS}$. Next, we apply $\mathtt{SortSched}$ (Algorithm \ref{alg:schedule}) to obtain a scheduling solution with respect to $\mathcal{L}_{H}$.
Since $\mathbf{z}$ is a feasible solution, Proposition \ref{prop:hjs1} applies. Thus, we first sort all schedule instances $\ell$ with positive $z_\ell$ as shown in lines $2$--$5$ of $\mathtt{SortSched}$. Note that we append the new schedule instance to the end of $\mathcal{F}$ for every while loop (line $4$), so the resulting set $\mathcal{F}$ is already sorted. Next, we show that we can always find such a schedule instance $\ell$ in line $3$ of $\mathtt{SortSched}$.

\begin{algorithm}[tb]
    \caption{Sort Scheduling ($\mathtt{SortSched}$)}\label{alg:schedule}
    \hspace*{\algorithmicindent} \textbf{Input}: $\mathbf{z}$ (an optimal fractional solution of $\mathsf{HIS}$) \\
    \hspace*{\algorithmicindent} \textbf{Output}: $\mathcal{S}_{H}$
    \begin{algorithmic}[1]
    \STATE  $\mathcal{F} \leftarrow \emptyset$. Remove all $\ell$ with $z_\ell = 0$ from $\mathcal{L}_{H}$;
    \WHILE{$\mathcal{L} \neq \emptyset$}
        \STATE Find a schedule instance $\ell$ that satisfies the inequality of Proposition \ref{prop:hjs1};
        \STATE Append $\ell$ to the end of $\mathcal{F}$, i.e., $\mathcal{F} \leftarrow \mathcal{F} \cup \{\ell\}$;
        \STATE Remove $\ell$ from $\mathcal{L}_{H}$, and set $z_\ell = 0$;
    \ENDWHILE
    \STATE $\mathcal{S}_{H} \leftarrow $ $\mathtt{FracLR}$($\mathcal{F}$, $\mathbf{e}$);
    \end{algorithmic}
\end{algorithm}

\begin{lemma}\label{lemma:hjs2}
    Let $\mathcal{F}$ be the resulting set of schedule instances from lines $2$--$5$ of $\mathtt{SortSched}$, and let $\ell^{k}$ denote the $k$-th schedule instance in $\mathcal{F}$. Let $\mathcal{F}[k] = \{\ell^{k}, \ell^{k+1}, ..., \ell^{|\mathcal{F}|}\}$. Then, we have 
    ${\textstyle z_{\ell^k} + \sum_{\ell' \in \mathcal{A}(\ell^k) \cap \mathcal{F}[k]}z_{\ell'} \le 6.}$
\end{lemma}
\begin{IEEEproof}
    By removing $z_{\ell^s}$ from $\mathbf{z}$ for $s < k$, the resulting $\mathbf{z}$ is still a feasible solution to $\mathsf{HIS}$. Thus, Proposition \ref{prop:hjs1} still applies to $\ell^k$ by only considering its neighbors in $\mathcal{F}[k]$.
\end{IEEEproof}

\begin{algorithm}[tb]
    \caption{Fractional Local Ratio ($\mathtt{FracLR}$)}\label{alg:flr}
    $\mathtt{FracLR}$($\mathcal{U}$, $\mathbf{w}$):
    \begin{algorithmic}[1]
    \STATE Remove all $\ell$ with non-positive energy $w(\ell)$ from $\mathcal{U}$.
    \STATE \textbf{if} $\mathcal{U} = \emptyset$, \textbf{return} $\emptyset$.
    \STATE Choose the $\ell$ with smallest index from $\mathcal{U}$. 
    \STATE Decompose the energy vector $\mathbf{w} = \mathbf{w_1} + \mathbf{w_2}$ such that $w_1(\ell') = w(\ell)$ if \scalebox{0.93}{$\ell' \in \mathcal{L}(\ell) \cup \mathcal{A}(\ell)$}; otherwise, \scalebox{0.93}{$w_1(\ell') = 0$}.
    \STATE $\mathcal{S}'_{F} \leftarrow $ $\mathtt{FracLR}$($\mathcal{U}$, $\mathbf{w_2}$).
    \STATE \textbf{if} $\mathcal{S}'_{F} \cup \{\ell\}$ is a feasible schedule, \textbf{return} $\mathcal{S}_{F} \triangleq \mathcal{S}'_{F} \cup \{\ell\}$; otherwise, \textbf{return} $\mathcal{S}_{F} \triangleq \mathcal{S}'_{F}$.
    \end{algorithmic}
\end{algorithm}

After we obtained the set of sorted schedule instance $\mathcal{F}$, we apply $\mathtt{FracLR}$ (Algorithm \ref{alg:flr}) to $\mathcal{F}$ to obtain a feasible scheduling solution for $\mathsf{EMJS}$ (line $6$ of $\mathtt{SortSched}$), where the $\mathbf{e}$ is the saved energy vector of all schedule instances. $\mathtt{FracLR}$ is a recursive algorithm, and in each recursive layer of $\mathtt{FracLR}$, we \textit{decompose the saved energy of schedule instances such that each schedule instance selection maintains a constant local approximation ratio corresponding to the optimal fractional solution $\mathbf{z}$ of $\mathsf{HIS}$}. Because the saved energy of each schedule instance is updated (decomposed) in each recursive layer, we use $w(\ell)$ to represent the updated saved energy of schedule instance $\ell$ in each layer ($e(\ell)$ denotes the original saved energy of $\ell$). The initial call to this recursive algorithm is $\mathtt{FracLR}$($\mathcal{F}$, $\mathbf{e}$). Next, we show that $\mathtt{SortSched}$ is a $7$-approximation algorithm for $\mathsf{HIS}$.

\begin{theorem}\label{theorem:hjs1}
    Suppose $\mathbf{z}$ is an optimal solution to $\mathsf{HIS}$, and $\mathcal{S}_{H}$ is the schedule returned by $\mathtt{FracLR}$ in line $6$ of $\mathtt{SortSched}$. Then, it holds that $e(\mathcal{S}_{H}) \ge \frac{1}{7}\cdot \mathbf{e} \cdot \mathbf{z}$.
\end{theorem}

\begin{IEEEproof}
    Let $w(\mathcal{S}_{F}) = \sum_{\ell \in \mathcal{S}_{F}} w(\ell)$. Note that any schedule instance removed by $\mathtt{FracLR}$ in step $1$ is considered to have zero weight. For ease of understanding, we denote the innermost recursive layer (when set $\mathcal{U} = \emptyset$) as layer $0$, and the outermost layer (initial call to $\mathtt{FracLR}$) as layer $U$. Besides, we use superscript $i$ to denote the parameters corresponding to recursive layer $i$, i.e., $\mathbf{w}^i$ denotes the energy vector $\mathbf{w}$ corresponding to recursive layer $i$. We prove this lemma by showing that $w^i(\mathcal{S}^i_{F}) \ge \frac{1}{7}\cdot \mathbf{w}^i \cdot \mathbf{z}$ for all $i = 0, 1, ..., U$. 
    
    In the base case ($i = 0$), $\mathcal{S}^0_{F} = \emptyset$ and the inductive hypothesis holds, since the weight vector $\mathbf{w}^0$ is considered to be zero. Next, we prove the inductive step.
    
    For $i \ge 1$, suppose $w^{i-1}(\mathcal{S}^{i-1}_{F}) \ge \frac{1}{7}\cdot \mathbf{w}^{i-1} \cdot \mathbf{z}$. According to line $4$ of $\mathtt{FracLR}$, $\mathbf{w}^{i}_\mathbf{2}$ is equivalent to $\mathbf{w}^{i-1}$ \textit{before} all schedule instances with non-positive energy are removed from $\mathcal{U}^{i-1}$. Adding non-positive component back to $\mathbf{w}^{i-1}$ will only decrease the value of $\mathbf{w}^{i}_\mathbf{2} \cdot \mathbf{z}$, thus, $w^{i-1}(\mathcal{S}^{i-1}_{F}) \ge \frac{1}{7} \cdot \mathbf{w}^{i-1} \cdot \mathbf{z} \ge \frac{1}{7} \cdot \mathbf{w}^{i}_\mathbf{2} \cdot \mathbf{z}$. Besides, in layer $i-1$, we have removed all schedule instances $\ell$ with nonpositive energy $w^{i-1}(\ell)$, and they will never be added to $\mathcal{S}^{i-1}_{F}$, thus, we have 
    \[w^{i}_2(\mathcal{S}'^{i}_{F})=w^{i}_2(\mathcal{S}^{i-1}_{F}) = w^{i-1}(\mathcal{S}^{i-1}_{F}).\]
    Since $w_1^i(\ell^i) = w^i(\ell^i)$, $w_2^i(\ell^i) = 0$; thus, $w^{i}_2(\mathcal{S}^{i}_{F}) = w^{i}_2(\mathcal{S}'^{i}_{F})$. Hence, $w^{i}_2(\mathcal{S}^{i}_{F}) = w^{i-1}(\mathcal{S}^{i-1}_{F}) \ge \frac{1}{7} \cdot \mathbf{w}^{i}_\mathbf{2} \cdot \mathbf{z}$.

    In $i$-th recursive layer of $\mathtt{FracLR}$, the chosen schedule instance $\ell^i$ has the smallest index in $\mathcal{U}^i$. According to Lemma \ref{lemma:hjs2} and step $3$ of $\mathtt{FracLR}$, $\sum_{\ell' \in \mathcal{A}(\ell) \cap \mathcal{U}^i} z_{\ell'} \le 6$. Due to constraint \eqref{eq:emjs3}, we have $\sum_{\ell' \in \mathcal{L}(\ell)} z_{\ell'} \le 1$. Thus, $\mathbf{w}^i_\mathbf{1} \cdot \mathbf{z} \le 7w^i(\ell)$. For the returned solution $\mathcal{S}_{F}$, it either contains schedule instance $\ell$ or contains at least one schedule instance in $\mathcal{L}(\ell) \cup \mathcal{A}(\ell)$. In both cases, $w_1^i(\mathcal{S}_{F})$ is at least $w^i(\ell)$. Thus, $w_1^i(\mathcal{S}^i_{F}) \ge \frac{1}{7}\mathbf{w}^i_\mathbf{1} \cdot \mathbf{z}$.
    
    Since the energy vector is decomposed in each layer such that $\mathbf{w}^i = \mathbf{w}^i_\mathbf{1} + \mathbf{w}_\mathbf{2}^i$, and the objective function is a linear multiplication of $\mathbf{w}$ and $\mathbf{z}$, we have 
    \[{\textstyle w^i(\mathcal{S}^i_{F}) = w^i_1(\mathcal{S}^i_{F}) + w^i_2(\mathcal{S}^i_{F}) \ge \frac{1}{7} \mathbf{w}^i_\mathbf{1} \cdot \mathbf{z} + \frac{1}{7} \mathbf{w}^i_\mathbf{2} \cdot \mathbf{z} = \mathbf{w}^i \cdot \mathbf{z}.}\]
    Based on the simple induction, we have $w^i(\mathcal{S}^i_{F}) \ge \frac{1}{7}\cdot \mathbf{w}^i \cdot \mathbf{z}$ for all $i = 0, 1, ..., U$. Since the initial call to $\mathtt{FracLR}$ (when $i = U$) is $\mathtt{FracLR}$($\mathcal{F}$, $\mathbf{e}$), the theorem is proved.
\end{IEEEproof}


\begin{theorem}\label{theorem:sjs}
    Let $\mathcal{S}$ be the scheduling solution obtained by $\mathtt{LHJS}$, and $\mathcal{S}^*$ be the optimal scheduling solution to $\mathsf{EMJS}$. Then, the expected value of $e(\mathcal{S})$ satisfies $\mathbf{E}[e(\mathcal{S})] \ge \frac{1}{23}e(\mathcal{S}^*)$.
\end{theorem}
\begin{IEEEproof}
    Let $OPT_L$ denote the optimal objective value of $\mathsf{LIS}$, and $OPT_H$ denote the optimal objective value of $\mathsf{HIS}$. According to Theorem \ref{theorem:ljs1} and Theorem \ref{theorem:hjs1}, we have $OPT_L \le 16\mathbf{E}[e(\mathcal{S}_{L})]$ and $OPT_H \le 7e(\mathcal{S}_{H})$. Furthermore, based on line $7$ of $\mathtt{LHJS}$, we have $e(\mathcal{S}) \ge \max\{e(\mathcal{S}_{L}), e(\mathcal{S}_{H})\}$.
    Since any feasible solution to $\mathsf{EMJS}$ can be divided into a set of light schedule instances and a set of heavy schedule instances, we have $ e(\mathcal{S}^*) \le OPT_L + OPT_H \le 16\mathbf{E}[e(\mathcal{S}_{L})] + 7e(\mathcal{S}_{H}) \le 23\mathbf{E}[e(\mathcal{S})]$. Therefore, $\mathbf{E}[e(\mathcal{S})] \ge \frac{1}{23}e(\mathcal{S}^*)$.
\end{IEEEproof}

\textbf{Discussion}. The partial elimination ordering technique employed in $\mathtt{SortSched}$ originates from Bar-Yehuda et al. \cite{bar2006split}, initially devised for scheduling jobs given multiple processing intervals on a single machine. Building upon their approach, we tackle the more complex sub-problem $(\mathsf{HIS})$: scheduling jobs with multiple operations, with multiple unrelated candidates (machines) for each operation. Because the total number of schedule instances depends on the total number of time units $\Delta$, $\mathtt{LHJS}$ is a pseudo-polynomial approximation algorithm for offline $\mathsf{EMJS}$. 
The offline scheduling algorithm can be used in applications where workloads are known apriori, such as intelligent patrol robots within factories or forest inspection drones that have regular working schedules, moving paths, and workloads (e.g., anomaly detection).

\begin{algorithm}[t]
    \caption{Load Balanced Job Scheduling ($\mathtt{LBS}$)}\label{alg:greedy}
    \hspace*{\algorithmicindent} \textbf{Input}: $\mathcal{N}(t_{cur})$, $\mathcal{M}$ \\
    \hspace*{\algorithmicindent} \textbf{Output}: $\mathcal{S}$
    \begin{algorithmic}[1]
    \STATE Initialize $\mathcal{S} \leftarrow \emptyset, \Omega \leftarrow \emptyset, $ and $\mathcal{N}^{sel} \leftarrow \emptyset$;
    \FORALL{$n_j \in \mathcal{N}(t_{cur}), T_u \in \psi_j, T_d \in \psi_j$}
    \STATE $\omega \leftarrow \langle n_j, T_u, T_d \rangle$;
    \IF{$ \exists t^u_j, t^d_j$ satisfying Eqs. \eqref{eq:sys1}, \eqref{eq:sys2}, \eqref{eq:sys5}, and \eqref{eq:sys6}}
    \STATE compute $e(\omega)$ as in Subsection \ref{subsec:archiB};
    \STATE Add $\omega$ into $\Omega$ if $e(\omega) > 0$;
    \ENDIF
    \ENDFOR
    \STATE sort $\Omega$ in non-increasing order of $e(\omega)$;
    \WHILE{$\Omega \neq \emptyset$}
    \STATE $\mathcal{U} \leftarrow \emptyset, \Omega \leftarrow \Omega \setminus \{\omega\}$;
    \STATE Let $\omega = \langle n_j,  T_{ur}, T_{ds} \rangle$ be the leftmost candidate in $\Omega$;
    \STATE \textbf{if} $n_j \in \mathcal{N}^{sel}$ \textbf{then} go to line $7$;
    \STATE Let $t^u_j$ be the earliest time such that vAP $m_u$ is idle during $[t^u_j, t^u_j+d_j^u-1]$ and $t^u_j$ satisfies Eqs. \eqref{eq:sys1}, \eqref{eq:sys2}; 
    \STATE Let $t^d_j$ be the latest time such that vAP $m_d$ is idle during $[t^d_j, t^d_j+d_j^d-1]$ and $t^d_j$ satisfies Eqs. \eqref{eq:sys5} and \eqref{eq:sys6};
    
    \FOR{$m_i \in \mathcal{M}^p_j$}
    \STATE Determine the earliest and latest times feasible for processing $t_{e}$ and $t_{l}$ based on $t^u_j$, $t^d_j$, Eqs. \eqref{eq:sys3}, \eqref{eq:sys4};

    $/\ast$ \scalebox{.83}[0.9]{\textit{let $\beta_i[t]$ be the used resource fraction of $m_i$ at time $t$}} $\ast/$
    \FOR{$c \in \mathcal{C}_i$ (from smallest to largest)}
    \IF{$\exists t'$ satisfying  $\beta_i[t] + c \le 1,$ $\forall t \in [t', t'+d_j^p-1]$, Eqs. \eqref{eq:sys3} and \eqref{eq:sys4}}
    \STATE $U_i = {(c \cdot d_j^p +\sum_{t = t_e}^{t_l}\beta_i[t])}/{(t_{l} -t_{e} +1)}$;
    \STATE $\mathcal{U} \leftarrow \mathcal{U} \cup \{U_i\}, c_{ij} \leftarrow c$;
    \STATE \textbf{break};
    \ENDIF
    \ENDFOR
    \ENDFOR

    \STATE \textbf{if} $\mathcal{U} = \emptyset$ \textbf{then} go to line $7$;
    \STATE $m_p \leftarrow \argmin_{m_i \in \mathcal{M}^p_j}U_i$;
    \STATE Given the resource allocation $c_{pj}$ (line $18$), let $t^p_j$ be the earliest time meeting the condition in line $16$ and has smallest $\scaleto{\max\{\beta_p[t] +c_{pj} \mid t \in [t^p_j, t^p_j+d_j^p-1]}{11.5pt}\}$;
    \STATE Update $t^d_j$ to the earliest time such that vAP $m_{ds}$ is idle during $[t^d_j, t^d_j+d_j^d-1]$ and $t^d_j$ satisfies Eqs. \eqref{eq:sys4}$\sim$\eqref{eq:sys6};
    \STATE Update resource allocation status for all $m_i \in \mathcal{M}$;
    \STATE \scalebox{.9}{$\mathcal{S}$$\leftarrow$$\mathcal{S} \cup \{\langle m_u, m_p, m_d, t^u_j, t^p_j, t^d_j, c_{pj}\rangle\}, \mathcal{N}^{sel}$$\leftarrow$$\mathcal{N}^{sel} \cup \{n_j\}$};
    \ENDWHILE
    \end{algorithmic}
\end{algorithm}

\section{Online Scheduling Algorithm for Online EMJS}\label{sec:heuristic}
In this section, we consider online $\mathsf{EMJS}$, where the scheduler only knows a job's information after its release in the system. In the online scenario, the scheduler is deployed on a central server within the backhaul network, while other servers and vAPs continuously update their resource utilization to the scheduler via the wired backhaul network. When jobs are generated, their meta-information is offloaded to the scheduler through accessible vAPs. Given that the size of job meta-information is typically very small, we omit its transmission time. The scheduler is triggered whenever a job arrives. To address online $\mathsf{EMJS}$, we introduce the Load Balanced Job Scheduling Algorithm ($\mathtt{LBS}$) invoked whenever a job is released in the system. The intuition behind $\mathtt{LBS}$ is to minimize the impact of the current job's scheduling on future job arrivals, such that more energy can be saved for future jobs. $\mathtt{LBS}$ is outlined in Algorithm \ref{alg:greedy}.

Let $\mathcal{N}(t_{cur})$ be the set of jobs whose meta-information arrives at the online scheduler at time $t_{cur}$. Let $\omega = \langle n_j, T_u, T_d\rangle$ be a ring coverage window selection for job $n_j$, where $T_u, T_d \in \psi_j$ are the selected offloading and downloading ring coverage windows, respectively. Since the energy saving of job $n_j$, $e_j^{save}$, depends on the selection of ring coverage windows, we first enumerate all possible window selection candidates $\omega$ for all jobs $n_j \in \mathcal{N}(t_{cur})$ and compute the corresponding saved energy, denoted as $e(\omega)$ (lines $2$--$5$).

Next, we consider the candidate $\omega = \langle n_j, T_{ur}, T_{ds})$ with the largest saved energy. We first determine the earliest starting time for offloading, $t^u_j$, and the latest starting time for downloading, $t^d_j$ (lines $8$--$11$). For each candidate server $m_i \in \mathcal{M}^p_j$, we determine job $n_j$'s feasible processing interval $[t_e, t_l]$, and then determine the smallest resource allocation option $c_{ij}$ such that $n_j$ can be processed within $[t_e, t_l]$ without violating server resource capacity constraint. If $c_{ij}$ can be found, we compute the resource usage (line $17$) of server $m_i$ during $[t_e, t_l]$ if job $n_j$ is processed on server $m$ with resource allocation $c_{ij}$ (lines $13$--$19$). After all candidate servers have been checked, we choose the server with the least resource usage as the processing server $m_p$ for job $n_j$. Then, we determine the earliest processing starting time, $t^p_j$, while minimizing the server's peak resource usage (line $22$). Finally, based on $t^p_j$, we update $t^d_j$ to the earliest downloading starting time. The algorithm stops when the candidate set, $\Omega$, becomes empty.

\section{Experimental Evaluation}\label{sec:experiment}
This section evaluates the performances of $\mathtt{LHJS}$ and $\mathtt{LBS}$. We first assess $\mathtt{LHJS}$ for offline job scheduling by comparing it with an offline algorithm in the literature. Then, we evaluate $\mathtt{LBS}$ for online job scheduling against its (three) variants. The experiments were conducted on a desktop PC equipped with an Intel(R) Xeon(R) W-2235 3.8GHz CPU and 32G RAM\footnote{Experiments code is available at \url{https://github.com/CPS-research-group/CPS-NTU-Public/tree/RTSS2024}.}.

\subsection{Simulation Setup}\label{subsec:exp-setup}
We select a city area of $1200m$$\times$$700m$ using OpenStreetMap \cite{OpenStreetMap}. Traffic data are generated using SUMO \cite{lopez2018microscopic}, simulating $675$ vehicles randomly entering and leaving this area within $600$ seconds. Each AP has a height of $30m$, and its network is divided into $2$ rings with coverage ranges of $0 \sim 100m$ and $100 \sim 200m$. We consider a MEC with $13$ APs, $6$ GPU servers, and $6$ CPU servers ($M^u$$=$$M^d$$=$$13$, $M^p$$=$$12$). The geographic coordinates of APs are determined with CellMapper \cite{cellmapper}, and servers are co-located with APs.


\begin{figure*}[t]
    \centering
    \subfigure[]{\includegraphics[width=0.32\textwidth, page=1]{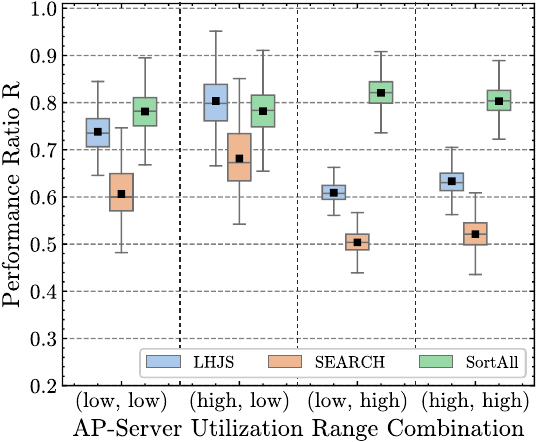} \label{fig:exp-result1}} 
    \subfigure[]{\includegraphics[width=0.32\textwidth, page=2]{figures/exp-results.pdf} \label{fig:exp-result2}}
    \subfigure[]{\includegraphics[width=0.32\textwidth]{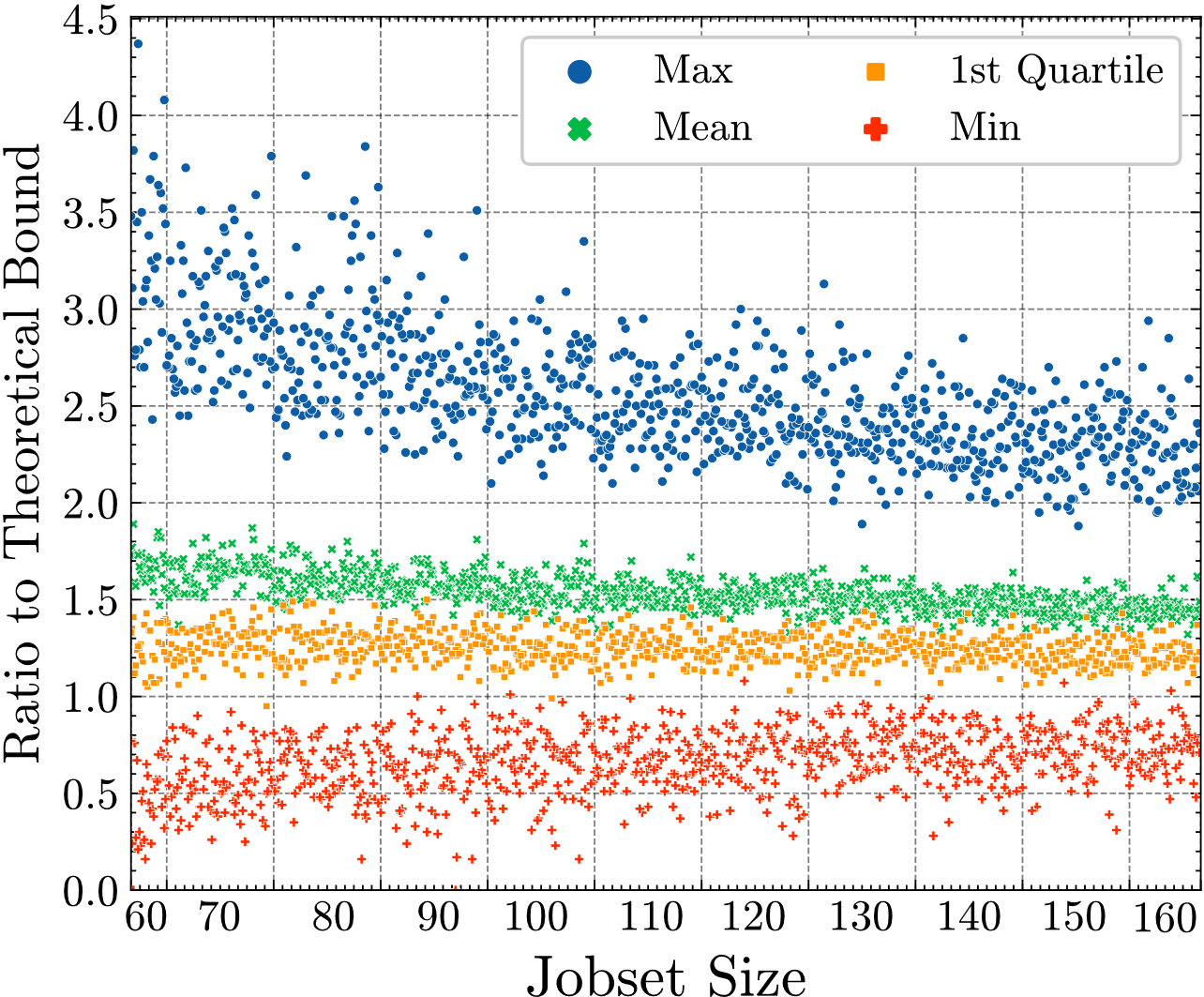} \label{fig:exp-result3}}
    \caption{(a) Performance ratios of $\mathtt{LHJS}$, $\mathtt{SEARCH}$, and $\mathtt{SortAll}$ under different AP-server utilization range combinations in offline job scheduling; (b) Intermediate results of $\mathtt{LHJS}$ compared to the optimal saved energy under different AP-server utilization range combinations. LIS-Round (likewise HIS-Round) is the integral solution for all light (likewise heavy) schedule instances via $\mathtt{RandRound}$ (likewise $\mathtt{SortSched}$); (c) A comparison between the integral solution obtained by $\mathtt{RandRound}$ and the approximation bound of $\mathtt{RandRound}$ (as in Theorem \ref{theorem:ljs1}).}
    \label{fig:exp1}
\end{figure*}

Nvidia Jetson Nano \cite{jetson_nano} is used as ED, and the wireless network data is obtained by profiling the wireless communication between Jetson Nano and a WiFi-5 router (TL-WDR8620). For each $m_i \in \mathcal{M}^u$, $\alpha_i$ is set to $40$ or $80$ MHz. When $\alpha_i=40$ (likewise $80$) MHz, the data offloading rate $\eta_j^u$ is set to $33$ and $23$ (likewise $66$ and $46.5$) MBps for ring $1$ and $2$, respectively. For each vAP $m_i \in \mathcal{M}^d$, the data downloading rate $\eta_j^d$ is set to $38$ and $77$ MBps for $\alpha_i=40$ and $\alpha_i=80$ MHz, respectively. We consider $3$ types of CPU (Intel w2235, 10700k and 14700k, with computing units $12, 16$, and $16$, respectively) and $3$ types of GPU (Nvidia 2080Ti, 4060Ti and 3090 with computing units $8, 8$, and $8$). Notably, with the same number of computing units, the processing time for the same job can differ with different server hardware.

We consider $3$ GPU applications (resnet101, resnet152, vgg-16) for object detection and $1$ CPU application (surf3D) for 3D-object surface reconstruction. The input $\theta_j^{in}$ of GPU applications includes $38$ images ranging from $0.01$ to $1.2$ MB, and that of surf3D includes $8$ 3D mesh objects ranging from $0.14$ to $0.6$ MB. For jobs of application resnet101, resnet152, vgg-16 and surf3D, their release times $\gamma_j$ are sampled from the ranges $[1, 100]$, $[1, 70]$, $[1, 90]$, and $[1, 50]$ milliseconds (ms), respectively, and their deadlines $\delta_j$ are set to $80$, $110$, $90$, and $130$ ms, respectively. The job (local and remote) processing time and ED's energy consumption for local processing, offloading, and downloading are obtained through profiling. Specifically, the power for processing CPU applications locally ranges from $0.97$ to $1.11$ Watts, and that for GPU applications ranges from $1.8$ to $5.33$ Watts. Besides, the average power of Jetson Nano for data offloading and downloading is $2.08$ and $2.13$ Watts, respectively.

In this experiment, we evaluate the algorithms' performance on jobsets with varying resource utilizations. The job deadlines are approximately $100$ ms; therefore, we use a scheduling window of $180$ ms for each jobset, since using a larger scheduling window does not necessarily lead to increased resource utilization of the jobset. The window starting times are randomly sampled from $120$ to $300$ seconds from the $600$ second simulation. We only consider the simulation period where most of the vehicles are active. Jobs are then randomly mapped to active vehicles during the scheduling window, obtaining ring coverage windows $\psi_j$ for each job by tracing the physical position of vehicles. Given a job mapping and computation resource allocation $c_j$ for job $n_j$, the computation resource utilization of job $n_j$ in a MEC is defined as $U^c_j = {c_j \cdot d_j^p}/({D \cdot d_{j}^{p,max}})$, where $d_{j}^{p,max}$ is the maximum allowable processing duration for job $n_j$ under the given mapping, and $D$ is the number of servers in the MEC. Considering various job mappings and $c_j$, let $U^{c,min}_{j}$ be the smallest utilization among all $U^c_j$ of job $n_j$. The total \textit{computation resource utilization} of a jobset is defined as $u^c = \sum_{j \in \mathcal{N}} U^{c,min}_{j}$, and the total uplink bandwidth utilization $u^b$ of a jobset is analogously defined. To generate jobsets with various computation and bandwidth resource utilization, we consider jobset sizes $J$ from a range of $[60, 160]$ (in increments of $10$). 
In total, we generate $3000$ jobsets with various utilizations.

\textbf{Offline Baseline.} For offline experiments, we compare $\mathtt{LHJS}$ with a heuristic algorithm ($\mathtt{SortAll}$) derived from $\mathtt{SortSched}$ and an approximation algorithm ($\mathtt{SEARCH}$) proposed by Zhu \emph{et al.} \cite{zhu2018task}. The key difference between $\mathtt{SortAll}$ and $\mathtt{SortSched}$ is that $\mathtt{SortAll}$ considers all schedule instances, rather than only heavy schedule instances. $\mathtt{SEARCH}$ first sorts all jobs in ascending order of deadlines and divides every consecutive $q$ jobs into a group; then, it applies an exhaustive search within each group to find the optimal schedule. Unlike our study, Zhu \emph{et al.} \cite{zhu2018task} did not consider (i) job waiting time for processing/downloading, (ii) varying computation resource allocation, or (iii) job mapping for processing/downloading. Therefore, $\mathtt{SEARCH}$ cannot be trivially extended to EMJS. However, the exhaustive search concept in $\mathtt{SEARCH}$ is applicable to any scheduling problem, including ours, making it a relevant baseline. To adapt $\mathtt{SEARCH}$ to $\mathsf{EMJS}$, we search for every possible job mapping and operation ordering within each group and set the group size $q=30$ to ensure that $\mathtt{SEARCH}$ has a runtime comparable to $\mathtt{LHJS}$. Notably, neither $\mathtt{SortAll}$ nor $\mathtt{SEARCH}$ provide an approximation guarantee for $\mathsf{EMJS}$.

\textbf{Online Baseline.} For online experiments, we compare $\mathtt{LBS}$ with its variants: $\mathtt{LBSLate}$, $\mathtt{LCEarly}$, and $\mathtt{LCLate}$. Compared with $\mathtt{LBS}$, $\mathtt{LBSLate}$ schedules each job's operation as late as possible, $\mathtt{LCEarly}$ considers only the largest computation resource allocation option for job processing, and $\mathtt{LCLate}$ considers only the largest computation resource allocation option and schedules each job's operation as late as possible.

\textbf{Metric}. We use the \textit{Performance Ratio} $R$ to measure the performance of all algorithms, where $R$ is defined as the ratio of the total saved energy by an algorithm to the optimal saved energy for $\mathsf{EMJS}$. Obtaining the optimal saved energy for  $\mathsf{EMJS}$ is challenging; hence, we utilize the optimal solution of a relaxed LP formulation of $\mathsf{EMJS}$, computed with the LP solver Gurobi (with a timeout limit of $10$ minutes), as an upper bound of the optimal to compute $R$. Hence, the computed $R$ is in fact a lower bound of the algorithm's actual performance.

\subsection{Offline Scheduling Experiment Result Discussion}

\textit{$\mathtt{LHJS}$ Performance.} Algorithms' performance are illustrated under different combinations of AP-Server utilization (i.e., ($u^b, u^c$)), where two different utilization ranges are employed for both APs and servers: \textbf{low} ($[0, 0.9)$) and \textbf{high} ($[1.1, \inf)$). The performances of $\mathtt{LHJS}$ $\mathtt{SEARCH}$, and $\mathtt{SortAll}$ under various utilization range combinations are shown in Fig. \ref{fig:exp-result1}, while insights into the intermediate results of $\mathtt{LHJS}$ are provided in Fig. \ref{fig:exp-result2}. On average, $\mathtt{LHJS}$, $\mathtt{SEARCH}$, and $\mathtt{SortAll}$ achieve performance ratios of $69.8\%$, $58\%$, and $79.5\%$, respectively. It is evident that $\mathtt{LHJS}$ consistently outperforms $\mathtt{SEARCH}$ across all utilization range combinations. Additionally, $\mathtt{SortAll}$ demonstrates superior performance when the computation resource utilization is high. While the performance of $\mathtt{LHJS}$ and $\mathtt{SEARCH}$ shows a monotonically increasing trend with AP utilization and a decreasing trend with server utilization, $\mathtt{SortAll}$ maintains stable performance across different utilization range combinations.


As shown in Fig. \ref{fig:exp-result2}, although the optimal fractional solution of $\mathsf{LIS}$ exceeds that of $\mathsf{HIS}$, the rounding loss induced by $\mathtt{RandRound}$ is significantly greater than $\mathtt{SortSched}$. Consequently, a larger integral solution is obtained for $\mathcal{L}_{H}$ compared to $\mathcal{L}_{L}$. Thus, the performance of $\mathtt{LHJS}$ is heavily influenced by the performance of $\mathtt{SortSched}$. Furthermore, since the rounding loss of $\mathtt{SortSched}$ is relatively small (around $2.3\%$), the practical performance of $\mathtt{LHJS}$ is primarily constrained by the optimal solution of $\mathsf{HIS}$. When the server utilization is low, the increasing AP utilization reduces the allowable time for job processing, favouring a large resource allocation option to shorten processing duration. Therefore, the optimal solution of $\mathsf{HIS}$ closely aligns with the optimal solution of $\mathsf{EMJS}$, leading to enhanced performance of $\mathtt{LHJS}$. When the server utilization increases from low to high, server resources become more scarce, and always allocating full server resources to a job may lead to a shortage of computing resources, resulting in a decreased optimal solution for $\mathsf{HIS}$. It is important to note that the objective value of $\mathsf{HIS}$ serves as an upper bound for all scheduling algorithms that assume full computation resource allocation. Considering the rounding loss of $\mathtt{SortSched}$ ($2.3\%$), $\mathtt{SortSched}$ offers a near-optimal practical solution for scenarios where fractional computation resource allocation is not permitted.

Based on Figs. \ref{fig:exp-result1} and \ref{fig:exp-result2}, we observe that the practical performance of $\mathtt{LHJS}$ is limited by only considering solutions from either light or heavy schedule instances to achieve a constant approximation ratio. In contrast, $\mathtt{SortAll}$ considers all schedule instances without aiming for theoretical guarantees, resulting in better practical performance compared to $\mathtt{LHJS}$. This also highlights that approximation algorithms can serve as a valuable foundation for designing heuristic algorithms with good practical performance.

\textit{$\mathtt{RandRound}$ Evaluation.} As $\mathtt{RandRound}$ is a randomized algorithm, we also evaluate its actual rounding performance. We conduct experiments with $900$ jobsets featuring various resource utilization combinations and sizes. For each jobset, utilizing the computed optimal fractional solution of $\mathsf{LIS}$, we execute $\mathtt{RandRound}$ $50$ times and record the resulting integral solutions. Among these $50$ integral solutions, we select $4$ results (maximum, minimum, average, and 1st quartile) and illustrate the ratios of these results to the approximation bound (Theorem \ref{theorem:ljs1}) in Fig. \ref{fig:exp-result3}. A higher ratio indicates better practical performance of $\mathtt{RandRound}$. The $1$st quartile denotes that $25\%$ of the $50$ integral solutions possess saved energies lower than the $1$st quartile. Observably, the ratio corresponding to the $1$st quartile of nearly all jobsets surpasses $1$, indicating a $75\%$ probability that the integral solution obtained by $\mathtt{RandRound}$ exceeds its theoretical bound. Additionally, the variance of the random rounding diminishes with increasing jobset size.

\begin{figure}
    \centering
    \includegraphics[width=0.9\linewidth, page=4]{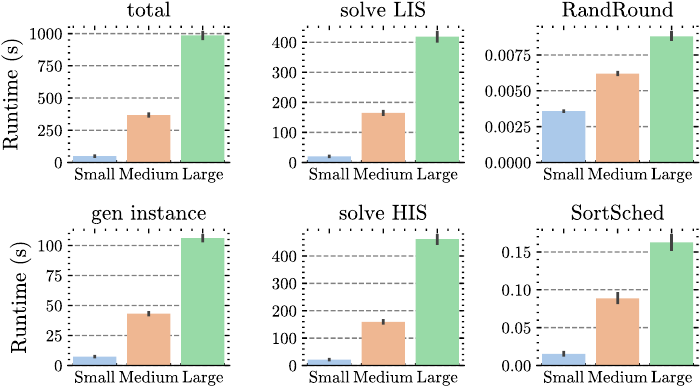}
    \caption{Runtime analysis of $\mathtt{LHJS}$ for different scales of MEC}
    \label{fig:exp-result4}
\end{figure}

\textit{$\mathtt{LHJS}$ Scalability.} To evaluate the scalability of $\mathtt{LHJS}$, we consider two additional MECs with larger scales: \textit{medium-scale} MEC (with $16$ APs and $18$ servers) and \textit{large-scale} MEC (with $21$ APs and $24$ servers). For each scale of MEC (including the base MEC which we now denote as \textit{small-scale} MEC), we generate $100$ jobsets with total resource utilizations $u^c$ and $u^b$ ranging from $0.9$ to $1.1$. Specifically, for small, medium, and large-scale MECs, each generated jobset contains $80$, $140$, and $200$ jobs, respectively. In Fig. \ref{fig:exp-result4}, we present the average runtimes of $\mathtt{LHJS}$ for different scales of MECs. The subfigure ``total'' represents the overall runtime of $\mathtt{LHJS}$, while the other five subfigures depict the runtime of individual steps of $\mathtt{LHJS}$ (i.e., ``gen instance'' denotes the time taken to define all schedule instances). It's evident that more than $10\%$ of the runtime is devoted to defining schedule instances, while over $85\%$ of the time is spent on solving the LP formulations $\mathsf{LIS}$ and $\mathsf{HIS}$. The time spent on LP rounding ($\mathtt{RandRound}$ and $\mathtt{SortSched}$) accounts for less than $0.1\%$ of the total runtime. Furthermore, the runtime of $\mathtt{LHJS}$ notably increases with the growing number of APs and servers in the MEC. This increase is primarily attributed to the significant rise in the number of schedule instances with the expansion of system scales and jobset sizes, posing a considerable challenge for the LP solver. 

\subsection{Online Scheduling Experiment Result Discussion}
\begin{figure}
    \centering
    \includegraphics[width=0.9\linewidth, page=5]{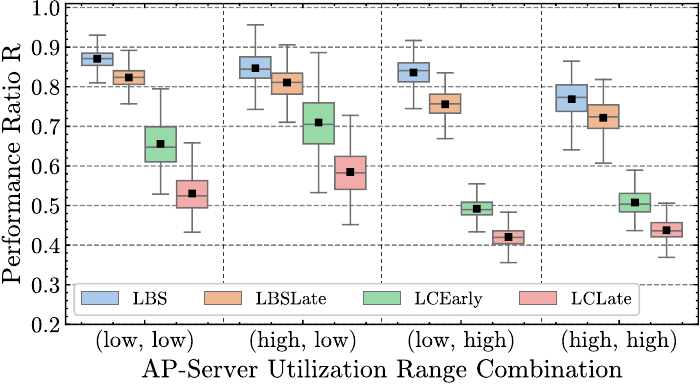}
    \caption{Performance ratios of $\mathtt{LBS}$ and its variants under different AP-server utilization range combinations for online job scheduling}
    \label{fig:exp-result5}
\end{figure}

\textit{$\mathtt{LBS}$ Performance.} In Fig. \ref{fig:exp-result5}, we present the performance of $\mathtt{LBS}$ and its variants under different AP-Server utilization range combinations for the base MEC (small-scale). On average, $\mathtt{LBS}$ achieves a performance ratio of $83.2\%$ compared to the offline optimal solution of $\mathsf{EMJS}$, surpassing its variants $\mathtt{LBSLate}$, $\mathtt{LCEarly}$, and $\mathtt{LCLate}$ by $5.2\%$, $23.7\%$, and $33.7\%$ respectively. Notably, the result highlights the significance of the computation resource allocation strategy in online scheduling compared to the policy for scheduling each job's operation (i.e., earliest or latest). Additionally, the performance gap between $\mathtt{LBS}$ and $\mathtt{LCEarly}$ widens as server utilization increases, which may stem from inefficient resource utilization with a large resource allocation option and the potentially significant impact of such an allocation on future job arrivals.

\begin{figure}
    \centering
    \includegraphics[width=0.9\linewidth, page=6]{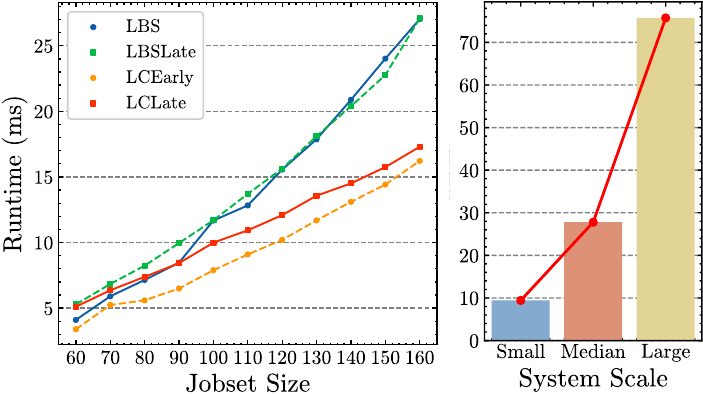}
    \caption{Runtime of $\mathtt{LBS}$ under different jobset sizes and MEC scales}
    \label{fig:exp-result6}
\end{figure}

\textit{$\mathtt{LBS}$ Scalability.} Fig. \ref{fig:exp-result6} illustrates the average runtime of $\mathtt{LBS}$ across different jobset sizes (in small-scale MEC) and varying MEC scales. It is evident that the $\mathtt{LBS}$ runtime increases almost linearly with the increase of jobset size. Despite considering various resource allocation options in $\mathtt{LBS}$, its runtime when compared to $\mathtt{LCEarly}$ and $\mathtt{LCLate}$ is only marginally higher (less than $15$ ms). Furthermore, the total runtime of $\mathtt{LBS}$ is merely $27$ ms for a jobset size of $160$. Considering the substantial performance improvement, this slight increase in runtime is justifiable. Moreover, even with a large-scale MEC and jobset size $200$, the runtime of $\mathtt{LBS}$ remains under $100$ ms, underscoring its practical viability.

\section{Conclusion}\label{sec:conclusion}
This paper addressed a deadline-constrained job mapping and resource management problem, $\mathsf{EMJS}$, in MEC with both communication and computation contentions, which jointly considered job scheduling, computation resource allocation, and ED mobility. For the offline $\mathsf{EMJS}$, we introduce a pseudo-polynomial approximation algorithm named $\mathtt{LHJS}$ with a constant approximation ratio. For the online $\mathsf{EMJS}$, we propose an online scheduling algorithm named $\mathtt{LBS}$ with practical runtimes. Experimental results show that both $\mathtt{LHJS}$ and $\mathtt{LBS}$ significantly outperform their baseline algorithms, and highlight that allocating full server resources to each job during processing may not be an effective strategy, particularly when the server resource demand of the jobset is high.

In this work, we considered a sequential data communication model (OFDM) in the wireless network. With the recent OFDMA data communication model in 5G, the subcarriers in each wireless channel can be assigned to different users simultaneously, leading to a similar resource allocation model as the server. In future work, we plan to explore approximation algorithms for job scheduling in the OFDMA-enabled MEC.

\bibliographystyle{IEEEtran}
\bibliography{IEEEabrv,reference}

\begin{thebibliography}{10}
\providecommand{\url}[1]{#1}
\csname url@samestyle\endcsname
\providecommand{\newblock}{\relax}
\providecommand{\bibinfo}[2]{#2}
\providecommand{\BIBentrySTDinterwordspacing}{\spaceskip=0pt\relax}
\providecommand{\BIBentryALTinterwordstretchfactor}{4}
\providecommand{\BIBentryALTinterwordspacing}{\spaceskip=\fontdimen2\font plus
\BIBentryALTinterwordstretchfactor\fontdimen3\font minus \fontdimen4\font\relax}
\providecommand{\BIBforeignlanguage}[2]{{%
\expandafter\ifx\csname l@#1\endcsname\relax
\typeout{** WARNING: IEEEtran.bst: No hyphenation pattern has been}%
\typeout{** loaded for the language `#1'. Using the pattern for}%
\typeout{** the default language instead.}%
\else
\language=\csname l@#1\endcsname
\fi
#2}}
\providecommand{\BIBdecl}{\relax}
\BIBdecl

\bibitem{ji2021guest}
W.~Ji, T.~Ebrahimi, Z.~Li, J.~Yuan, D.~O. Wu, and Y.~Xin, ``Guest editorial: Emerging visual iot technologies for future communications and networks,'' \emph{IEEE Wireless Communications}, vol.~28, no.~4, pp. 10--11, 2021.

\bibitem{aazam2018fog}
M.~Aazam, S.~Zeadally, and K.~A. Harras, ``Fog computing architecture, evaluation, and future research directions,'' \emph{IEEE Communications Magazine}, vol.~56, no.~5, pp. 46--52, 2018.

\bibitem{schneider2017augmented}
M.~Schneider, J.~Rambach, and D.~Stricker, ``Augmented reality based on edge computing using the example of remote live support,'' in \emph{2017 IEEE International Conference on Industrial Technology (ICIT)}.\hskip 1em plus 0.5em minus 0.4em\relax IEEE, 2017, pp. 1277--1282.

\bibitem{soyata2012cloud}
T.~Soyata, R.~Muraleedharan, C.~Funai, M.~Kwon, and W.~Heinzelman, ``Cloud-vision: Real-time face recognition using a mobile-cloudlet-cloud acceleration architecture,'' in \emph{2012 IEEE symposium on computers and communications (ISCC)}.\hskip 1em plus 0.5em minus 0.4em\relax IEEE, 2012, pp. 000\,059--000\,066.

\bibitem{yurtsever2020survey}
E.~Yurtsever, J.~Lambert, A.~Carballo, and K.~Takeda, ``A survey of autonomous driving: Common practices and emerging technologies,'' \emph{IEEE access}, vol.~8, pp. 58\,443--58\,469, 2020.

\bibitem{ma2011battery}
C.~Ma and Y.~Yang, ``A battery-aware scheme for routing in wireless ad hoc networks,'' \emph{IEEE Transactions on Vehicular Technology}, vol.~60, no.~8, pp. 3919--3932, 2011.

\bibitem{shuai2022transfer}
K.~Shuai, Y.~Miao, K.~Hwang, and Z.~Li, ``Transfer reinforcement learning for adaptive task offloading over distributed edge clouds,'' \emph{IEEE Transactions on Cloud Computing}, 2022.

\bibitem{ma2024latency}
B.~Ma, Z.~Ren, W.~Cheng, J.~Wang, and W.~Zhang, ``Latency-constrained multi-user efficient task scheduling in large-scale internet of vehicles,'' \emph{IEEE Transactions on Mobile Computing}, 2024.

\bibitem{zhang2024computational}
X.~Zhang, T.~Lin, C.-K. Lin, Z.~Chen, and H.~Cheng, ``Computational task offloading algorithm based on deep reinforcement learning and multi-task dependency,'' \emph{Theoretical Computer Science}, p. 114462, 2024.

\bibitem{alameddine2019dynamic}
H.~A. Alameddine, S.~Sharafeddine, S.~Sebbah, S.~Ayoubi, and C.~Assi, ``Dynamic task offloading and scheduling for low-latency iot services in multi-access edge computing,'' \emph{IEEE Journal on Selected Areas in Communications}, vol.~37, no.~3, pp. 668--682, 2019.

\bibitem{sorkhoh2019workload}
I.~Sorkhoh, D.~Ebrahimi, R.~Atallah, and C.~Assi, ``Workload scheduling in vehicular networks with edge cloud capabilities,'' \emph{IEEE Transactions on Vehicular Technology}, vol.~68, no.~9, pp. 8472--8486, 2019.

\bibitem{xu2023adaptive}
Y.~Xu, L.~Chen, Z.~Lu, X.~Du, J.~Wu, and P.~C.~K. Hung, ``An adaptive mechanism for dynamically collaborative computing power and task scheduling in edge environment,'' \emph{IEEE Internet of Things Journal}, vol.~10, no.~4, pp. 3118--3129, 2023.

\bibitem{lou2022cost}
J.~Lou, Z.~Tang, S.~Zhang, W.~Jia, W.~Zhao, and J.~Li, ``Cost-effective scheduling for dependent tasks with tight deadline constraints in mobile edge computing,'' \emph{IEEE Transactions on Mobile Computing}, 2022.

\bibitem{gao2023joint}
Y.~Gao, J.~Tao, H.~Wang, Z.~Wang, W.~Sun, and C.~Song, ``Joint server deployment and task scheduling for the maximal profit in mobile edge computing,'' \emph{IEEE Internet of Things Journal}, 2023.

\bibitem{abdi2024task}
S.~Abdi, M.~Ashjaei, and S.~Mubeen, ``Task offloading in edge-cloud computing using a q-learning algorithm,'' in \emph{The International Conference on Cloud Computing and Services Science, CLOSER (2024)}, 2024.

\bibitem{sang2024mobility}
Y.~Sang, J.~Wei, Z.~Zhang, and B.~Wang, ``A mobility-aware task scheduling by hybrid pso and ga for mobile edge computing,'' \emph{Cluster Computing}, pp. 1--16, 2024.

\bibitem{tiwari2024rate}
M.~Tiwari, I.~Maity, and S.~Misra, ``Rate: Reliability-aware task service in fog-enabled iov environments,'' \emph{IEEE Transactions on Cognitive Communications and Networking}, 2024.

\bibitem{zhu2018task}
T.~Zhu, T.~Shi, J.~Li, Z.~Cai, and X.~Zhou, ``Task scheduling in deadline-aware mobile edge computing systems,'' \emph{IEEE Internet of Things Journal}, vol.~6, no.~3, pp. 4854--4866, 2019.

\bibitem{huang2023joint}
X.~Huang and G.~Huang, ``Joint optimization of energy and task scheduling in wireless-powered irs-assisted mobile-edge computing systems,'' \emph{IEEE Internet of Things Journal}, vol.~10, no.~12, pp. 10\,997--11\,013, 2023.

\bibitem{huang2024mobility}
L.~Huang and Q.~Yu, ``Mobility-aware and energy-efficient offloading for mobile edge computing in cellular networks,'' \emph{Ad Hoc Networks}, vol. 158, p. 103472, 2024.

\bibitem{meng2019dedas}
J.~Meng, H.~Tan, C.~Xu, W.~Cao, L.~Liu, and B.~Li, ``Dedas: Online task dispatching and scheduling with bandwidth constraint in edge computing,'' in \emph{IEEE INFOCOM 2019-IEEE Conference on Computer Communications}.\hskip 1em plus 0.5em minus 0.4em\relax IEEE, 2019, pp. 2287--2295.

\bibitem{han2020joint}
D.~Han, W.~Chen, and Y.~Fang, ``Joint channel and queue aware scheduling for latency sensitive mobile edge computing with power constraints,'' \emph{IEEE Transactions on Wireless Communications}, vol.~19, no.~6, pp. 3938--3951, 2020.

\bibitem{pu2019chimera}
L.~Pu, X.~Chen, G.~Mao, Q.~Xie, and J.~Xu, ``Chimera: An energy-efficient and deadline-aware hybrid edge computing framework for vehicular crowdsensing applications,'' \emph{IEEE Internet of Things Journal}, vol.~6, no.~1, pp. 84--99, 2019.

\bibitem{pradhan2024towards}
S.~Pradhan, S.~Tripathy, and R.~Matam, ``Towards optimal edge resource utilization: Predictive analytics and reinforcement learning for task offloading,'' \emph{Internet of Things}, p. 101147, 2024.

\bibitem{chauhan2024probabilistic}
N.~Chauhan and R.~Agrawal, ``A probabilistic deadline-aware application offloading in a multi-queueing fog system: A max entropy framework,'' \emph{Journal of Grid Computing}, vol.~22, no.~1, pp. 1--22, 2024.

\bibitem{li2024uav}
W.~Li, S.~Li, H.~Shi, W.~Yan, and Y.~Zhou, ``Uav-enabled fair offloading for mec networks: a drl approach based on actor-critic parallel architecture,'' \emph{Applied Intelligence}, pp. 1--18, 2024.

\bibitem{lai2024short}
X.~Lai, T.~Wu, C.~Pan, L.~Mai, and A.~Nallanathan, ``Short-packet edge computing networks with execution uncertainty,'' \emph{IEEE Transactions on Green Communications and Networking}, 2024.

\bibitem{ben2020maximizing}
L.~Ben~Yamin, J.~Li, K.~Sarpatwar, B.~Schieber, and H.~Shachnai, ``Maximizing throughput in flow shop real-time scheduling,'' \emph{Leibniz international proceedings in informatics}, vol. 176, no.~48, 2020.

\bibitem{jamil2022resource}
B.~Jamil, H.~Ijaz, M.~Shojafar, K.~Munir, and R.~Buyya, ``Resource allocation and task scheduling in fog computing and internet of everything environments: A taxonomy, review, and future directions,'' \emph{ACM Computing Surveys (CSUR)}, vol.~54, no. 11s, pp. 1--38, 2022.

\bibitem{luo2021resource}
Q.~Luo, S.~Hu, C.~Li, G.~Li, and W.~Shi, ``Resource scheduling in edge computing: A survey,'' \emph{IEEE Communications Surveys \& Tutorials}, vol.~23, no.~4, pp. 2131--2165, 2021.

\bibitem{ramanathan2020survey}
S.~Ramanathan, N.~Shivaraman, S.~Suryasekaran, A.~Easwaran, E.~Borde, and S.~Steinhorst, ``A survey on time-sensitive resource allocation in the cloud continuum,'' \emph{it-Information Technology}, vol.~62, no. 5-6, pp. 241--255, 2020.

\bibitem{intel_wifi_protocal}
\BIBentryALTinterwordspacing
Intel, ``Different wi-fi protocols and data rates,'' 2023. [Online]. Available: \url{https://www.intel.com/content/www/us/en/support/articles/000005725/wireless/legacy-intel-wireless-products.html}
\BIBentrySTDinterwordspacing

\bibitem{hamdi2021tran}
M.~M. Hamdi and M.~S. Abood, ``Transmission over ofdm and sc-fdma for lte systems,'' in \emph{Intelligent Systems Design and Applications}, A.~Abraham, V.~Piuri, N.~Gandhi, P.~Siarry, A.~Kaklauskas, and A.~Madureira, Eds.\hskip 1em plus 0.5em minus 0.4em\relax Cham: Springer International Publishing, 2021, pp. 722--731.

\bibitem{ankarali2020enhanced}
Z.~E. Ankaral{\i}, B.~Pek{\"o}z, and H.~Arslan, ``Enhanced ofdm for 5g ran,'' \emph{ZTE Communications}, vol.~15, no.~S1, pp. 11--20, 2020.

\bibitem{li2006orthogonal}
Y.~G. Li and G.~L. Stuber, \emph{Orthogonal frequency division multiplexing for wireless communications}.\hskip 1em plus 0.5em minus 0.4em\relax Springer Science \& Business Media, 2006.

\bibitem{zhou2022online}
R.~Zhou, J.~Pang, Q.~Zhang, C.~Wu, L.~Jiao, Y.~Zhong, and Z.~Li, ``Online scheduling algorithm for heterogeneous distributed machine learning jobs,'' \emph{IEEE Transactions on Cloud Computing}, 2022.

\bibitem{fang2020constgat}
\BIBentryALTinterwordspacing
X.~Fang, J.~Huang, F.~Wang, L.~Zeng, H.~Liang, and H.~Wang, ``Constgat: Contextual spatial-temporal graph attention network for travel time estimation at baidu maps,'' in \emph{Proceedings of the 26th ACM SIGKDD International Conference on Knowledge Discovery \& Data Mining}, ser. KDD '20.\hskip 1em plus 0.5em minus 0.4em\relax New York, NY, USA: Association for Computing Machinery, 2020, p. 2697–2705. [Online]. Available: \url{https://doi.org/10.1145/3394486.3403320}
\BIBentrySTDinterwordspacing

\bibitem{xia2013commercial}
T.~J. Xia and G.~A. Wellbrock, ``Commercial 100-gbit/s coherent transmission systems,'' \emph{Optical Fiber Telecommunications}, pp. 45--82, 2013.

\bibitem{shannon1984}
C.~E. Shannon, ``A mathematical theory of communication,'' \emph{The Bell System Technical Journal}, vol.~27, no.~3, pp. 379--423, 1948.

\bibitem{wang2006realistic}
Q.~Wang, M.~Hempstead, and W.~Yang, ``A realistic power consumption model for wireless sensor network devices,'' in \emph{2006 3rd Annual IEEE Communications Society on Sensor and Ad Hoc Communications and Networks}, vol.~1, 2006, pp. 286--295.

\bibitem{vaidya1987an}
\BIBentryALTinterwordspacing
P.~M. Vaidya, ``An algorithm for linear programming which requires o(((m+n)n2+(m+n)1.5n)l) arithmetic operations,'' in \emph{Proceedings of the Nineteenth Annual ACM Symposium on Theory of Computing}, ser. STOC '87.\hskip 1em plus 0.5em minus 0.4em\relax New York, NY, USA: Association for Computing Machinery, 1987, p. 29–38. [Online]. Available: \url{https://doi.org/10.1145/28395.28399}
\BIBentrySTDinterwordspacing

\bibitem{bar2006split}
\BIBentryALTinterwordspacing
R.~Bar-Yehuda, M.~M. Halld\'{o}rsson, J.~S. Naor, H.~Shachnai, and I.~Shapira, ``Scheduling split intervals,'' \emph{SIAM Journal on Computing}, vol.~36, no.~1, pp. 1--15, 2006. [Online]. Available: \url{https://doi.org/10.1137/S0097539703437843}
\BIBentrySTDinterwordspacing

\bibitem{OpenStreetMap}
\BIBentryALTinterwordspacing
O.~contributors, ``Planet dump retrieved from https://planet.osm.org,'' 2017. [Online]. Available: \url{https://www.openstreetmap.org}
\BIBentrySTDinterwordspacing

\bibitem{lopez2018microscopic}
P.~A. Lopez, M.~Behrisch, L.~Bieker-Walz, J.~Erdmann, Y.-P. Fl{\"o}tter{\"o}d, R.~Hilbrich, L.~L{\"u}cken, J.~Rummel, P.~Wagner, and E.~Wie{\ss}ner, ``Microscopic traffic simulation using sumo,'' in \emph{2018 21st international conference on intelligent transportation systems (ITSC)}.\hskip 1em plus 0.5em minus 0.4em\relax IEEE, 2018, pp. 2575--2582.

\bibitem{cellmapper}
\BIBentryALTinterwordspacing
``Cellmapper.'' [Online]. Available: \url{https://www.cellmapper.net/map}
\BIBentrySTDinterwordspacing

\bibitem{jetson_nano}
\BIBentryALTinterwordspacing
Nvidia, ``Nvidia jetson nano,'' 2024. [Online]. Available: \url{https://www.nvidia.com/en-sg/autonomous-machines/embedded-systems/jetson-nano/product-development/}
\BIBentrySTDinterwordspacing

\end{thebibliography}

\end{document}